\documentclass[iop,apj]{emulateapj}

\usepackage[usenames,dvips]{color}
\usepackage{graphicx}

\newcommand{\Msun}{\,M_\odot}
\newcommand{\Rsun}{\,R_\odot}
\newcommand{\Lsun}{\,L_\odot}

\newcommand{\gtap}{\gtrsim}
\newcommand{\ltap}{\lesssim}

\begin{document}

\title{Population Synthesis of Common Envelope Mergers: \\I. Giant Stars with Stellar or Substellar Companions}

\author{Michael Politano}
\affil{Department of Physics, Marquette University, P.O.\ Box 1881, Milwaukee, WI 53201-1881}

\author{Marc van der Sluys\altaffilmark{1,2}}
\affil{Department of Physics, University of Alberta, 11322 - 89 Ave, Edmonton AB, T6G 2G7, Canada}
\altaffiltext{1}{CITA National Fellow}
\altaffiltext{2}{Department of Physics and Astronomy, Northwestern University, 2131 Tech Drive, Evanston, IL 60208}

\and

\author{Ronald E.\ Taam\altaffilmark{3}, and Bart Willems}
\affil{Department of Physics and Astronomy, Northwestern University, 2131 Tech Drive, Evanston, IL 60208}
\altaffiltext{3}{Academia Sinica Institute of Astronomy \& Astrophysics/TIARA, Taipei, Taiwan}

\shorttitle{Common-envelope mergers on the giant branches}
\shortauthors{Politano, van der Sluys, Taam \& Willems}

\begin{abstract}

Using a population synthesis technique,  we have calculated detailed models of the present-day field population of objects 
that have resulted from the merger
of a giant primary and a main-sequence or brown dwarf secondary during common-envelope evolution.
We used a grid of 116 stellar and 32 low-mass/brown dwarf models, a crude model of the merger process, 
and followed the angular momentum evolution of the binary orbit and the primary's rotation prior to merger, 
as well as the merged object's rotation after the merger.  We find that present-day merged objects that are observable as 
giant stars or core-helium burning stars in our model population constitute between 0.24\% and 0.33\% of the initial population of 
ZAMS binaries, depending upon the input parameters chosen.
The median projected rotational velocity of these merged objects is $\sim 16$\,km\,sec$^{-1}$, 
an order of magnitude higher than the median projected rotational velocity in a model population of normal single stars calculated using the same
stellar models and initial mass function.  The masses of the merged objects are typically less than $\sim 2\Msun$, 
with a median mass of $1.28\Msun$, which is slightly more than, but not
significantly different from, their normal single star counterparts.  The luminosities in our merged object population 
range from $\sim 10 - 100\Lsun$, with a strong peak in the luminosity distribution at $\sim 60\Lsun$, since the 
majority of the merged objects (57\%) lie on the horizontal branch
at the present epoch.  The results of our population synthesis study 
are discussed in terms of possible observational counterparts  
either directly involving the high rotational velocity of the merger product or indirectly, via the effect of rotation on  
envelope abundances and on the amount and distribution of circumstellar matter.  

\end{abstract}


\section{Introduction}
\label{sec:intro}

There is a general consensus that the common-envelope (CE) phase plays an essential role in the evolution 
of many close binary stars.  During CE evolution, one star becomes engulfed in the envelope of its companion.  
The orbit of the binary decays due to the action of gravitational torques \citep[\textit{e.g.},][]{rick08} 
between it and the non-co-rotating envelope.  CE evolution involving stars in various evolutionary states can 
occur, but the ultimate fate of such evolution, regardless of the states of the component stars, is either a 
merger or a stable binary system \citep[see][for reviews]{ibe93,taa00}.  

There has been a renewed interest in CE mergers, specifically in massive binaries where merger of the 
cores or of a core and in-spiraling compact object has been suggested as a mechanism for producing 
hypernovae and gamma-ray bursts \citep[\textit{e.g.},][]{fry98,mid04,fry05}. In this paper, we investigate the 
evolutionary consequences of CE mergers involving lower-mass binaries consisting of a low- to intermediate-mass 
giant star and a stellar or substellar companion.  Research on CE evolution involving low-mass binaries has 
focused almost exclusively on systems that survive CE evolution.  In part, this is due to the identification 
of observational counterparts to such surviving systems (\textit{e.g.}, cataclysmic variables, double degenerate 
white dwarfs) and the lack of such obvious counterparts to mergers. However, detailed hydrodynamical 
calculations of CE evolution, incomplete as they are, indicate that mergers should occur in such systems 
\citep[\textit{e.g.},][]{san98,san00}.  Furthermore, population synthesis of close binary stars that involve a single phase of 
CE evolution between a low- to intermediate-mass giant star and a stellar or substellar companion predict that 
mergers should be as frequent as post-CE binaries 
in the disk of our Galaxy \citep{pw07}. Such studies suggest that there may exist a significant number of 
merged objects in our Galaxy that have yet to be recognized as such observationally.  
 
Only a handful of studies of the consequences of CE mergers in low-mass binaries exist in the literature. 
\citet{pod01} provides an overview of some of the physical considerations involved in modeling CE mergers in general.  
\citet{sie99a,sie99b}, \citet{sok00, sok07}, and \citet{car09} investigated the spiral-in and evaporation of 
planets or brown dwarfs (BDs) inside the envelope of a giant star. These authors found that the result of such a 
merger is a rapidly rotating single star, and they suggest that such mergers could explain certain abundance 
anomalies, such as enhanced Li, observed in the envelopes of a small fraction of giant stars, as well as the 
heterogeneity found among horizontal branch (HB) stars.  
Blue stragglers, single subdwarf B (sdB) or subdwarf O (sdO) stars, and R Coronae Borealis stars, to name a few, have been suggested 
as possible observational counterparts to low-mass CE mergers \citep[\textit{e.g.},][]{tut05,sok07,pol08}.  
However, there have been only two population synthesis studies 
that explored the consequences of low-mass CE mergers: \citet{izz07}, who modeled the population of early-type
R stars, and \citet{pol08}, who modeled the population of 
single sdB stars.

In this paper, we extend our previous work \citep{pol08} and present population synthesis calculations of the 
present-day population of objects in the disk of our Galaxy that have resulted from mergers during a single 
phase of CE evolution involving a giant primary and a stellar or substellar secondary.  To the best of our 
knowledge, these are the first population synthesis calculations that comprehensively explore the consequences of 
mergers that occur during CE evolution involving low-mass binaries.  In \S\,\ref{sec:method}, we describe the method used to model 
this population and discuss the assumptions underlying our study. In \S\,\ref{sec:results}, the results of our calculations 
are presented and compared to a corresponding present-day population of normal single stars calculated using the 
same grid of stellar-evolution models.  In \S\,\ref{sec:discussion}, we discuss the observational signatures of CE mergers and 
speculate about potential observational counterparts to CE mergers.  Finally, we summarize and conclude 
in the last section (\S\,\ref{sec:conclusions}).

\section{Method}
\label{sec:method}

\subsection{Population synthesis code}
\label{sec:code}

We use the Monte-Carlo population synthesis code developed by Politano \citep{pol96,pw07} with two major modifications:
(1)  the analytic fits that had been used previously (see \citealp{pol96}) have been replaced by numerical tables of 
116 up-to-date stellar models ranging in mass from 0.5 to $10.0\Msun$ in increments of $0.1\Msun$ and from 10.5 to 
$20\Msun$ in increments of $0.5\Msun$ and (2) tidal effects which act to synchronize the rotation of the primary with 
the orbital motion of the system have been included.  

The updated stellar models were calculated using a version of the binary stellar evolution code \texttt{ev}\footnote{The 
current version of \texttt{ev} is obtainable on request from \texttt{eggleton1@llnl.gov}, along with data files and a user 
manual.} developed by Eggleton \citep[][and references therein]{1971MNRAS.151..351E, 1972MNRAS.156..361E,2005ApJ...629..1055Y} 
and updated as described 
in \citet{1995MNRAS.274..964P}.
Convective mixing is modelled by a diffusion equation, using a
mixing-length to pressure scale-height ratio $l/H_\mathrm{p} = 2.0$.  Convective overshooting in the core is taken into account
on the main sequence for stars with $M \gtap 1.2\,M_\odot$ and on the horizontal branch for stars of all masses.
We use an overshooting parameter $\delta_\mathrm{ov}=0.12$, which corresponds to an overshooting length of about 
$0.3\,H_\mathrm{p}$.  The code cannot evolve a model through the helium flash, but this is resolved by automatically
replacing the model at the moment of degenerate helium ignition by a tailored model with the same total mass and core mass in which
helium was just ignited non-degenerately.  For stars with $M \gtap 2.1\,M_\odot$, 
helium ignites non-degenerately and this intervention is not needed.
The helium-core mass $M_\mathrm{c}$ is defined as the mass coordinate at which the hydrogen abundance drops below 10\%.
We compute the binding energy of the hydrogen-rich envelope, $E_\mathrm{bind}$, by integrating the gravitational and internal
energies over the mass coordinate of the model, from the core-envelope boundary to the surface of the star.
In the internal energy we include the thermal energy, but not the recombination energy.
More details regarding these assumptions are provided in \cite{2006A&A...460..209V}.  
The initial composition of our model stars is $X=0.70$, $Y=0.28$ and $Z=0.02$.
Mass loss via stellar winds is incorporated in the models using a prescription that was originally inspired by \citet{rei75}:
\begin{equation}
  \dot{M} \! = \! -\eta \times \min \! \left\{ \begin{array}{ll}
      \!3.16\!\times\!10^{-14}M_\odot\,\mathrm{yr}^{-1} \left(\!\frac{M}{M_\odot}\!\right) \!\! \left(\!\frac{L}{L_\odot}\!\right) \!\! \left(\!\frac{E_\mathrm{bind}}{10^{50} \mathrm{erg}}\!\right)^{\!-1}  \\
      \!9.61\!\times\!10^{-10}M_\odot\,\mathrm{yr}^{-1} \left(\!\frac{L}{L_\odot}\!\right) 
    \end{array}\!\right.\!\!,
  \label{eq:wind} 
\end{equation}

where we use $\eta = 0.2$ in this study, and the prescription 
by \cite{1988A&AS...72..259D}, which is negligible except for the most massive stars in our grid.
The upper prescription in Eq.\,\ref{eq:wind} is used in the majority of our models, except where the 
envelope binding energy is small (in absolute value), \textit{e.g.}, for stars near the tip of the asymptotic giant branch (AGB).

As a second modification, we now assume synchronism of the primary's rotation with the orbital motion whenever the 
primary is on either giant branch.  In these cases, after each time step, the 
new total angular momentum of the system is calculated, accounting for losses due to stellar winds.  
Synchronism is 
then enforced by redistributing this total angular momentum, $J_\mathrm{tot}$, such that the rotational angular velocity of the primary is 
equal to the orbital angular velocity of the binary. The new orbital separation is calculated as the root of the 
following equation:
\begin{equation}
  z^4 - b z^3 + c = 0.
  \label{eq:amloss}
\end{equation}
In this equation, $z = (a/R)^{1/2}$, where $a$ is the orbital separation and $R$ is the primary's radius, and $b$ and 
$c$ are dimensionless constants given by $b = J_\mathrm{tot}/[G\mu^2(M_\mathrm{p}+M_\mathrm{s})R]^{1/2}$ and 
$c = k^2(1+M_\mathrm{p}/M_\mathrm{s})$,
where $G$ is the universal gravitation constant, $M_\mathrm{p}$ and $M_\mathrm{s}$ are the primary and secondary masses, respectively,
$\mu$ is the reduced mass of the binary, and $k$ is the radius of gyration of the primary.  In all cases, 
the rotational angular momentum of the secondary is neglected and a 
circular orbit and rigid-body rotation for the primary are assumed.

In our population synthesis calculations, we begin with $10^7$ zero-age main sequence (ZAMS) binaries.  We assume 
that these binaries are distributed over primary mass ($M_\mathrm{p}$) according to a \citet{mil79} initial-mass 
function (IMF), over orbital 
period uniformly in $\log P$ \citep{abt83} and over mass ratios uniformly in $q$ (\textit{i.e.}, $g(q)\,dq = 1\,dq$, 
\citealp{duq91,maz92,gol03}), where $q = M_\mathrm{s}/M_\mathrm{p}$. 
Given the large observational uncertainties in the distribution of mass ratios in ZAMS binaries, we also investigate 
the dependence of our results on the assumed choice for $g(q)$ (see \S\,\ref{sec:input}).
We adopt a minimum primary mass of $0.95\Msun$, a maximum primary mass of $10\Msun$ and a minimum secondary mass of 
$0.013\Msun$ in our calculations.  For secondaries with masses less than $0.5\Msun$, including substellar secondaries, 
we use detailed stellar models from the Lyon group \citep{cb97,bar98,bar03,cha00}.  For secondaries with masses 
greater than or equal to $0.5\Msun$, we use the same stellar models as for the primary. We assume that the CE phase 
is so rapid that its duration is negligible compared to the other time scales \citep[\textit{e.g.},][]{ibe93,taa00}, 
the age of the Galaxy is $10^{10}$\,yrs (representative of the thin-disk population), and the star formation 
rate throughout the Galaxy's history has remained constant.  Lastly, we do not take into account angular momentum 
loss due to magnetic braking.

\subsection{Merger scenario}
\label{sec:mergerscenario}

To model the population of mergers between giant primaries and less massive companions, we adopt the following 
scenario. A given ZAMS binary is evolved until the primary reaches the base of the red-giant branch (RGB), at which point synchronism 
between the primary's rotation and the orbit is assumed.  As the primary ascends the RGB, two conditions are checked at the end of each time step:
(1) has the primary filled its Roche lobe or (2) has a tidal instability developed (see below).  
If the Roche lobe has been filled, we check to see if the mass transfer will be unstable by using the criterion given in \citet{hur02} (their Eq.\,57): 
\begin{equation}
  q_\mathrm{crit} = \frac{2.13}{1.67 - x + 2(M_\mathrm{c}/M_\mathrm{p})^5},
  \label{eq:qcrit}
\end{equation} 
\noindent where $M_\mathrm{p}$ and $M_\mathrm{c}$ are the total mass and core mass of the primary when it fills its 
Roche lobe and $x = 0.30$ if  $M_\mathrm{c} \leq 0.50\Msun$ or $x = 0.33$ if  $M_\mathrm{c} > 0.50\Msun$.  
If $q < q_\mathrm{crit}$, the mass transfer will be unstable and a CE phase will ensue.  
If $q >  q_\mathrm{crit}$, the mass transfer will be stable and the system will avoid a CE phase.  
The evolution of such systems is no longer followed and these systems are removed from the 
population\footnote{We recognize that such systems may potentially undergo a CE phase later in their evolution 
(\textit{e.g.}, if unstable mass transfer is initiated when the secondary fills its Roche lobe).  
However, our code does not treat phases of stable mass transfer.}. 
If a tidal instability has developed, we assume a CE phase will inevitably ensue.

If the primary does not fill its Roche lobe or if a tidal instability does not develop on the RGB, 
the primary is evolved through the core He-burning phase and then resynchronized 
at the base of the AGB.  
As the primary ascends the AGB,  the same two conditions as before are checked at the end of each time step and the same procedure is followed.
If neither condition is satisfied when the primary 
reaches the tip of the AGB, the system is removed from the population.  

To determine if merger occurs within the CE, simple energy considerations are used to relate the pre- and post-CE 
orbital separations \citep{tut79}.  Although the exact prescription varies somewhat depending upon the author, a 
typical expression used in population synthesis calculations is given below 
\citep[\textit{e.g.},][]{ibe85,pol96,wil05} 
\begin{equation}
  -\alpha_\mathrm{CE}\left(\frac{GM_\mathrm{c}M_\mathrm{s}}{2a_\mathrm{f}} - \frac{GM_\mathrm{p}M_\mathrm{s}}{2a_\mathrm{i}}\right)
  = E_\mathrm{bind,i},
  \label{eq:ubind}
\end{equation}
\noindent where 
$a_\mathrm{i}$ and $a_\mathrm{f}$ are the pre- and post-CE orbital 
separations, respectively, $E_\mathrm{bind,i}$ is the binding energy of the primary's envelope at the onset of the CE phase,
and $\alpha_\mathrm{CE}$ is a parameter that measures the efficiency with which orbital energy is transferred to 
the CE. The parameter $\alpha_\mathrm{CE}$ embodies a major uncertainty in this simplified prescription.  We have 
chosen $\alpha_\mathrm{CE} = 1$ as our standard model and we investigate the dependence of our results on 
$\alpha_\mathrm{CE}$ (see \S\,\ref{sec:input}).  
The term $E_\mathrm{bind,i}$ includes both the gravitational binding energy and thermal internal energy of 
primary's envelope, and is computed directly from the detailed stellar-structure models.  We note that this eliminates the need to parametrize the 
envelope's binding energy using a dimensionless constant, $\lambda$ (\textit{i.e.}, $E_\mathrm{bind,i} = 
GM_\mathrm{e}[M_\mathrm{e} + M_\mathrm{c}]/\lambda R_\mathrm{i}$, where 
$M_\mathrm{e}$ and $R_\mathrm{i}$ are the envelope mass and radius of the primary at the onset of the CE phase). 
Merger is assumed to have occurred if the radius of the secondary is larger than 
its Roche-lobe radius at the end of the CE phase. 

We also consider mergers that result from a tidal instability.  If the moment of inertia of the primary 
exceeds one-third of the moment of inertia of the binary orbit before the primary fills its Roche lobe 
on the RGB or AGB, 
Eq.\,\ref{eq:amloss} has no real roots. Physically, this implies that there will no longer be sufficient angular momentum 
within the orbit to keep the envelope rotating synchronously \citep{dar79} and the orbit will continue to shrink 
until mass transfer starts, after which either the envelope is ejected or a merger occurs.

\subsubsection{Treatment of the merger process}
\label{sec:treatment}

Available detailed hydrodynamical models of stellar mergers have focused on collisions between stars 
and are more appropriate for mergers that occur as a result of stellar interactions within dense stellar environments, 
such as at the centers of globular clusters \citep[\textit{e.g.},][]{sil97,sil01,lom02}.  In the absence of detailed models that 
are applicable to stellar mergers during CE evolution, a very simplified model for the merger process is adopted. 
The key assumptions underlying our treatment are listed below. 

\begin{enumerate}
\item In most cases, the merger of part or all of the secondary with the primary leads to an object spinning at 
several times its break-up rotational velocity, given by $v_\mathrm{br} = (GM/R)^{1/2}$, where $M$ and $R$ are 
the object's mass and radius, respectively.  Consequently, we assume that a rapid mass-loss phase occurs when the 
rotational velocity of the object exceeds some critical rotational velocity, $v_\mathrm{crit} = \beta\,v_\mathrm{br}$.  
For our standard model, we assume $\beta= 1/3$, which corresponds to centrifugal forces contributing about 
10\% of the force against gravity.  Our choice is motivated by the fact that we have adopted mass loss rates for 
single stars, including giant stars, that are slowly rotating.  At higher rotational velocities, it is likely that strong 
magnetic activity would be present, significantly enhancing the mass loss rates.  The rate of mass loss via such 
a process is not well understood, and we have adopted standard rates to avoid introducing an additional parameter in 
this preliminary study. To explore the qualitative effect of the choice of $\beta$ on our results, we calculated model populations 
for other values of $\beta$ (see \S\,\ref{sec:input}). 
 
\item The merged object is assumed to assimilate only as much mass from the binary as is required such that it 
rotates at the critical velocity specified under 1, and we assume that the remaining mass is expelled from the system.  
We note that the latter is a simplifying assumption.  In reality, it may be that only a portion of this remaining mass 
actually has enough energy to escape from the system, with the rest eventually accreted by the 
merged object (likely through an accretion disk).

\item The material in the merged object is assumed to be grossly distributed according to its specific entropy, 
with lower-entropy material residing in the core and higher-entropy material residing in the envelope.  Consequently, 
we assume that  the merged object has a core mass that is equal to that of the original primary, 
but an envelope mass that is potentially larger than that of the primary.

\item Although the merged object is unlikely to be in a state of thermal equilibrium initially, we assume for 
definiteness that the radius of the merged object corresponds to that of a star in the model grid for the same 
core mass and total mass after it achieves thermal equilibrium. Such an approximation may underestimate the
amount of angular momentum that the remnant may acquire immediately after the merger process, which could lead to
an increased level of magnetic activity due to dynamo action and an elevated level of mass and angular momentum loss 
(see item 1 above).

\item In some cases, the merger results in an object with a total mass/core mass combination for which no 
corresponding giant-branch model exists in our stellar grid.  In the majority of these cases, the core mass of 
the merged object is less than the core mass at the base of the corresponding giant branch for a stellar model 
with the same total mass.  Rather than permit an unphysical evolutionary regression, such as an AGB primary 
reverting to a core He-burning star as a result of the merger, we choose the stellar model in our grid at the 
base of the corresponding giant branch with the same total mass as the merged object.  Clearly this is not optimal, 
but the difference in core mass is generally small and, at present, we can do no better.  

\item If the CE takes place when the primary is on the RGB, the merger may result in a total-mass/core-mass 
combination that corresponds to a core-He burning star in our model grid (i.e., since the critical core mass for degenerate
He ignition decreases with increasing total mass, the additional envelope material acquired during merger 
may reduce the critical core mass needed for He ignition).  Since it is 
physically reasonable for an RGB star to ignite helium in the core, even if induced by a merger event,
we adopt a core He-burning model for the merged object. 
\end{enumerate}

\subsection{Evolution of the merged object to the present epoch} 
\label{sec:evolution}

The subsequent evolution of the merged object, including its rotational velocity, $v_\mathrm{rot}$, is followed 
until the present epoch using the same grid of stellar evolution models that we used for the primary prior to merger. 
In addition to angular momentum loss due to stellar winds, we assume that angular momentum is removed from the merged object's envelope
via rotationally enhanced mass loss whenever $v_\mathrm{rot}$ exceeds $v_\mathrm{crit}$.  
Typically, this occurs when there is a significant decrease in the merged object's radius, such as during contraction to the HB following He ignition. 
If $v_\mathrm{rot} > v_\mathrm{crit}$ during a given time step, we calculate the amount of angular momentum that needs to be lost so 
that the merged object will rotate exactly at $v_\mathrm{crit}$ and we artificially remove the corresponding amount of mass from the envelope. 
Once $0.1\Msun$ of material is lost, it becomes necessary to shift to 
the next lowest-mass model in our grid with the same core mass in order to continue to follow 
the object's evolution.  If there is no 
corresponding lower-mass model with the same core mass, we follow the procedure described in assumption 5 in the previous section.  However, in 
some cases this procedure proved unsatisfactory and led to unphysical behavior.  For example, some objects would 
continue to lose mass, 
moving along the base of the HB until the object moved off the grid.  To address this, we softened our condition for shifting to the next 
lowest-mass model by letting the object evolve for five additional time steps before shifting.  
We found that if the object's rotational velocity dropped below critical 
during these five time steps, its rotational velocity stabilized and 
the evolution could generally be followed for the remainder of its evolution
without problem.
If the merged object completes its evolution as an AGB star before the present epoch is reached, we assume 
it will be a white dwarf at the present epoch, and we remove the object from the population.

\subsection{Input parameters}
\label{sec:input}

For our standard model, we have assumed values for the CE efficiency parameter and ZAMS
mass ratio distribution that are common in the literature, namely, $\alpha_\mathrm{CE} = 1$ and $g(q)\,dq = 1\,dq$. 
In addition, we have chosen a value of $\beta = 1/3$ in our standard model, where the parameter $\beta$ represents the 
fraction of the break-up velocity for which we assume rotationally-induce mass transfer begins  
(see item 1 of \S\,\ref{sec:treatment}).  Given the large uncertainties in these parameters, we have calculated 
model populations for other choices of these parameters.  
We list the parameter choices made for each model in Table~\ref{tab:models}.

\begin{deluxetable}{lccc}
  \tablecolumns{4}
  \tablecaption{
    Input parameter choices for the models in this study
    \label{tab:models}
  }
  \tablewidth{0pc}
  \tablehead{
   \colhead{Model} & \colhead{$\alpha_\mathrm{CE}$} & \colhead{$g(q)$}  & \colhead{$\beta$} \\
  }
  \startdata
  standard & 1.0 & 1 & $\onethird$ \\
  $\alpha_\mathrm{CE}$ = 0.5                    &     0.5     &  1                  &  $\onethird$  \\
  $\alpha_\mathrm{CE}$ = 0.1                    &     0.1     &  1                  &  $\onethird$  \\
  $g(q) = q$                                                     &     1.0     &  $q$             & $\onethird$   \\
  $g(q) = q^{-0.9}$                                         &     1.0     &  $q^{-0.9}$  &  $\onethird$  \\
  $v_\mathrm{crit} = v_\mathrm{br}$          &     1.0     &  1                  &  1.0                 \\
  $v_\mathrm{crit} = 0.1\,v_\mathrm{br}$  &     1.0     &  1                  &  0.1                 \\

 \enddata
\end{deluxetable}

\section{Results}
\label{sec:results}

In this section, we present the results of our population synthesis calculations.  In \S\,\ref{sec:properties}, we discuss 
the general characteristics of the present-day population of merged objects, with an emphasis on observable 
quantities such as rotational velocity, luminosity, effective temperature, etc.  All results in \S\,\ref{sec:properties} 
are for our standard model (see Table~\ref{tab:models}).
In \S\,\ref{sec:dependence}, we discuss the dependence of our results on other choices for these three parameters,
as listed in Table~\ref{tab:models}.
Finally, in \S\,\ref{sec:single}, we compare our model population of merged objects with a standard single-star population 
calculated from the same set of stellar evolutionary models.

\subsection{Properties of present-day population of merged objects}
\label{sec:properties}

\subsubsection{Relative numbers and sub-populations}
\label{sec:numbers}

In our population synthesis calculations, we consider a population of $10^7$ primordial binaries with 
initial properties as described in \S\,\ref{sec:code}.  In our standard model, $\sim 71\%$ of these
binaries never initiate mass transfer, 13\% undergo stable mass transfer and 16\% undergo unstable mass transfer, which
leads to a CE phase.  Of those binaries undergoing CE evolution, 48\% survive the CE phase, while 52\% merge
(8.5\% of the original population).  
82\% of the CEs are formed when
the primary fills its Roche lobe, whereas 18\% undergo a Darwin instability.
Furthermore, 87\% of the mergers occur when the primary is on the RGB, while 13\% occur when the primary is on the AGB.

In the evolution following merger, the vast majority (97\%) of the merged objects will have evolved beyond the AGB by the 
present epoch.  These merger remnants would be observed as white dwarfs today and are not included in our present-day population of 
merged objects.  Only 3\% of the 
merged objects ($\sim 25,000$ stars or 0.25\% of our original population) would be observed as 
non-degenerate stars today.  The properties of these 25,000 merged objects are listed in Table~\ref{tab:dependence}.

\begin{deluxetable*}{lccccccccc}
  \tablecolumns{10}
  \tablecaption{
    Dependence of selected model results upon input parameters
    \label{tab:dependence}
  }
  \tablewidth{0pc}
  \tablehead{
    \colhead{Model} & \colhead{$N$\tablenotemark{a}} & \colhead{$\frac{N}{N_\mathrm{tot}}$\tablenotemark{b}} & \colhead{$M$\tablenotemark{c}} & 
    \colhead{$v \sin i$\tablenotemark{d}} &  \multicolumn{2}{c}{Fraction with\tablenotemark{e}} & $M_\mathrm{rej}$\tablenotemark{f} & 
    $\frac{M_\mathrm{rej}}{M_\mathrm{bin}}$\tablenotemark{g} & $\frac{\Delta M_\mathrm{mrg}}{M_\mathrm{mrg,i}}$\tablenotemark{h}\\
    \cline{6-7}
    \colhead{} & \colhead{} & \colhead{} & \colhead{($M_\odot$)} & \colhead{(km/s)} & \colhead{$v_\mathrm{rot}\!\leq\!0.1\,v_\mathrm{crit}$} & 
    \colhead{$v_\mathrm{rot}\!=\!v_\mathrm{crit}$} & \colhead{($M_\odot$)} & \colhead{} & \colhead{}
  } 
  \startdata
  
  \cutinhead{RGB stars}
  standard                                &     9301  &  0.37  &  1.20  &      18.4  &  (0.001)  &  0.0319  &        0.63  &   0.34  &   0.00  \\
  $\alpha_\mathrm{CE}$ = 0.5              &     9534  &  0.34  &  1.20  &      17.7  &  (0.0009)  &  0.0343  &        0.64  &   0.34  &   0.00  \\
  $\alpha_\mathrm{CE}$ = 0.1              &     9634  &  0.29  &  1.20  &      17.4  &  (0.0009)  &  0.0363  &        0.64  &   0.34  &   0.00  \\
  $g(q) = q$                              &     9044  &  0.36  &  1.20  &      18.2  &  (0.0009)  &  0.0368  &        0.89  &   0.42  &   0.00  \\
  $g(q) = q^{-0.9}$                       &     8819  &  0.35  &  1.20  &      18.4  &  (0.0007)  &  0.0251  &        0.16  &   0.11  &   0.00  \\
  $v_\mathrm{crit} = v_\mathrm{br}$       &     8025  &  0.33  &  1.30  &      48.7  &  (0.0006)  &  0.0179  &        0.45  &   0.25  &   0.00  \\
  $v_\mathrm{crit} = 0.1\,v_\mathrm{br}$  &     9783  &  0.38  &  1.19  &       5.4  & (0.001)  &  0.0250  &        0.71  &   0.37  &   0.00  \\
  single stars                            &   178651  &  0.61  &  1.20  &       1.9  &  0.9627  &  0.000  &       \ldots & \ldots  & \ldots  \\
  
  \cutinhead{HB stars}
  standard                                &    14305  &  0.57  &  1.35  &      16.1  &  (0.0000)  &  0.3149  &        0.93  &   0.40  &   0.12  \\
  $\alpha_\mathrm{CE}$ = 0.5              &    16987  &  0.60  &  1.34  &      16.7  &  (0.0000)  &  0.3469  &        0.91  &   0.41  &   0.12  \\
  $\alpha_\mathrm{CE}$ = 0.1              &    21043  &  0.64  &  1.23  &      17.3  &  (0.0000)  &  0.4091  &        0.91  &   0.42  &   0.13  \\
  $g(q) = q$                              &    14267  &  0.57  &  1.35  &      16.1  &  (0.0000)  &  0.3275  &        1.25  &   0.49  &   0.12  \\
  $g(q) = q^{-0.9}$                       &    14929  &  0.59  &  1.35  &      16.5  &  (0.0000)  &  0.3264  &        0.32  &   0.20  &   0.11  \\
  $v_\mathrm{crit} = v_\mathrm{br}$       &    14943  &  0.61  &  1.58  &      50.5  &  (0.0004)  &  0.2098  &        0.74  &   0.32  &   0.08  \\
  $v_\mathrm{crit} = 0.1\,v_\mathrm{br}$  &    14217  &  0.56  &  1.24  &       4.5  &  (0.0008)  &  0.3366  &        1.03  &   0.44  &   0.12  \\
  single stars                            &   104979  &  0.36  &  1.58  &       3.2  &  0.0886  &  0.0021  &       \ldots & \ldots  & \ldots  \\
  
  \cutinhead{AGB stars}
  standard                                &     1435  &  0.06  &  1.34  &       6.0  &  0.0683  &  (0.0007)  &        0.94  &   0.42  &   0.13  \\
  $\alpha_\mathrm{CE}$ = 0.5              &     1748  &  0.06  &  1.31  &       6.3  &  0.0732  &  (0.0011)  &        0.94  &   0.42  &   0.13  \\
  $\alpha_\mathrm{CE}$ = 0.1              &     2205  &  0.07  &  1.23  &       6.4  &  0.0757  &  (0.0018)  &        0.92  &   0.43  &   0.14  \\ 
  $g(q) = q$                              &     1542  &  0.06  &  1.34  &       6.4  &  0.0746  &  (0.0013)  &        1.27  &   0.49  &   0.13  \\
  $g(q) = q^{-0.9}$                       &     1595  &  0.06  &  1.34  &       6.0  &  0.0678  &  (0.0031)  &        0.36  &   0.23  &   0.12  \\
  $v_\mathrm{crit} = v_\mathrm{br}$       &     1446  &  0.06  &  1.56  &      18.7  &  0.0788  &  (0.0007)  &        0.74  &   0.33  &   0.09  \\ 
  $v_\mathrm{crit} = 0.1\,v_\mathrm{br}$  &     1490  &  0.06  &  1.23  &       1.8  &  0.0859  &  (0.0007)  &        1.09  &   0.46  &   0.14  \\
  single stars                            &    10487  &  0.04  &  1.45  &       1.3  &  0.5657  &  (0.0000)  &       \ldots & \ldots  & \ldots  \\
  
  \cutinhead{Total population}
  standard                                &    25041  &  1.00  &  1.28  &      16.2  &  0.0043  &  0.1918  &        0.81  &   0.38  &   0.07  \\
  $\alpha_\mathrm{CE}$ = 0.5              &    28269  &  1.00  &  1.23  &      16.2  &  0.0048  &  0.2201  &        0.81  &   0.38  &   0.08  \\
  $\alpha_\mathrm{CE}$ = 0.1              &    32882  &  1.00  &  1.23  &      16.5  &  0.0054  &  0.2726  &        0.83  &   0.40  &   0.10  \\
  $g(q) = q$                              &    24853  &  1.00  &  1.29  &      16.0  &  0.0049  &  0.2015  &        1.10  &   0.46  &   0.07  \\
  $g(q) = q^{-0.9}$                       &    25343  &  1.00  &  1.29  &      16.3  &  0.0045  &  0.2012  &        0.28  &   0.18  &   0.07  \\
  $v_\mathrm{crit} = v_\mathrm{br}$       &    24414  &  1.00  &  1.46  &      47.7  &  0.0051  &  0.1343  &        0.63  &   0.30  &   0.02  \\
  $v_\mathrm{crit} = 0.1\,v_\mathrm{br}$  &    25490  &  1.00  &  1.20  &       4.6  &  0.0058  &  0.1974  &        0.90  &   0.41  &   0.08  \\
  single stars                            &   294117  &  1.00  &  1.23  &       2.3  &  0.6366  &  0.0008  &       \ldots & \ldots  & \ldots  \\
  \enddata
  \tablenotetext{a}{total number of merged objects in present-day population}
  \tablenotetext{b}{fraction of present-day population in this evolutionary state}
  \tablenotetext{c}{median mass of merged objects in present-day population}
  \tablenotetext{d}{median projected rotational velocity of merged objects in present-day population}
  \tablenotetext{e}{fraction of merged objects in present-day population rotating at one-tenth of the critical velocity 
    or less and at the critical velocity (defined in \S\,\ref{sec:treatment})}
  \tablenotetext{f}{median amount of mass that is not accreted by the remnant during merger}
  \tablenotetext{g}{same as f, except given as a fraction of the total mass of the binary at the onset of the CE phase}
  \tablenotetext{h}{median of the relative difference between the mass immediately after the merger event, $M_\mathrm{mrg,i}$, 
    and the mass at the present epoch, $M_\mathrm{mrg,PE}$:
    $\frac{M_\mathrm{mrg,i}-M_\mathrm{mrg,PE}}{M_\mathrm{mrg,i}}$} 
  \tablecomments{
    Entries enclosed by parentheses indicate the number of objects is so small as to not be statistically significant.
  }
\end{deluxetable*}

We have divided the total population of 25,000 present-day merged objects into three sub-populations: RGB stars, 
HB stars and AGB stars. The properties of each of these sub-populations are also listed in Table~\ref{tab:dependence}.  
We find that HB stars are the most numerous in our present-day population (57\%), followed by RGB stars (37\%) 
and then AGB stars (6\%). 
This can be understood intuitively from the fact that 
of the merger remnants observed today, 99.9\% were formed in a merger on the RGB, compared to 0.1\% for which the 
merger took place on the AGB.  In particular, we find that mergers during CE evolution 
typically occur in binaries with primary masses of $\sim 1-2\Msun$ that fill their Roche lobes on the lower half of
the RGB at orbital periods between roughly 1 and 30 days.  The distribution of mass ratios at the onset of the CE phase is almost uniform
between 0 and 1.
While RGB mergers produce remnants that must evolve through the rest of the RGB phase, the relatively long-lasting HB phase 
and the AGB phase, AGB mergers produce remnants that evolve off the AGB quickly and disappear from our present-day population.
Of all the RGB mergers, only 2.4\% 
of the remnants ignite helium in the core and become HB stars as an immediate consequence of the merger.  Most of 
the remaining RGB mergers will have evolved off the RGB by the time of the present epoch, and the distribution of 
stars over the three branches is to a large extent determined by the relative lifetimes of the stars in each phase.

\subsubsection{Rotational velocity} 
\label{sec:vrot}

\begin{figure}
  \plotone{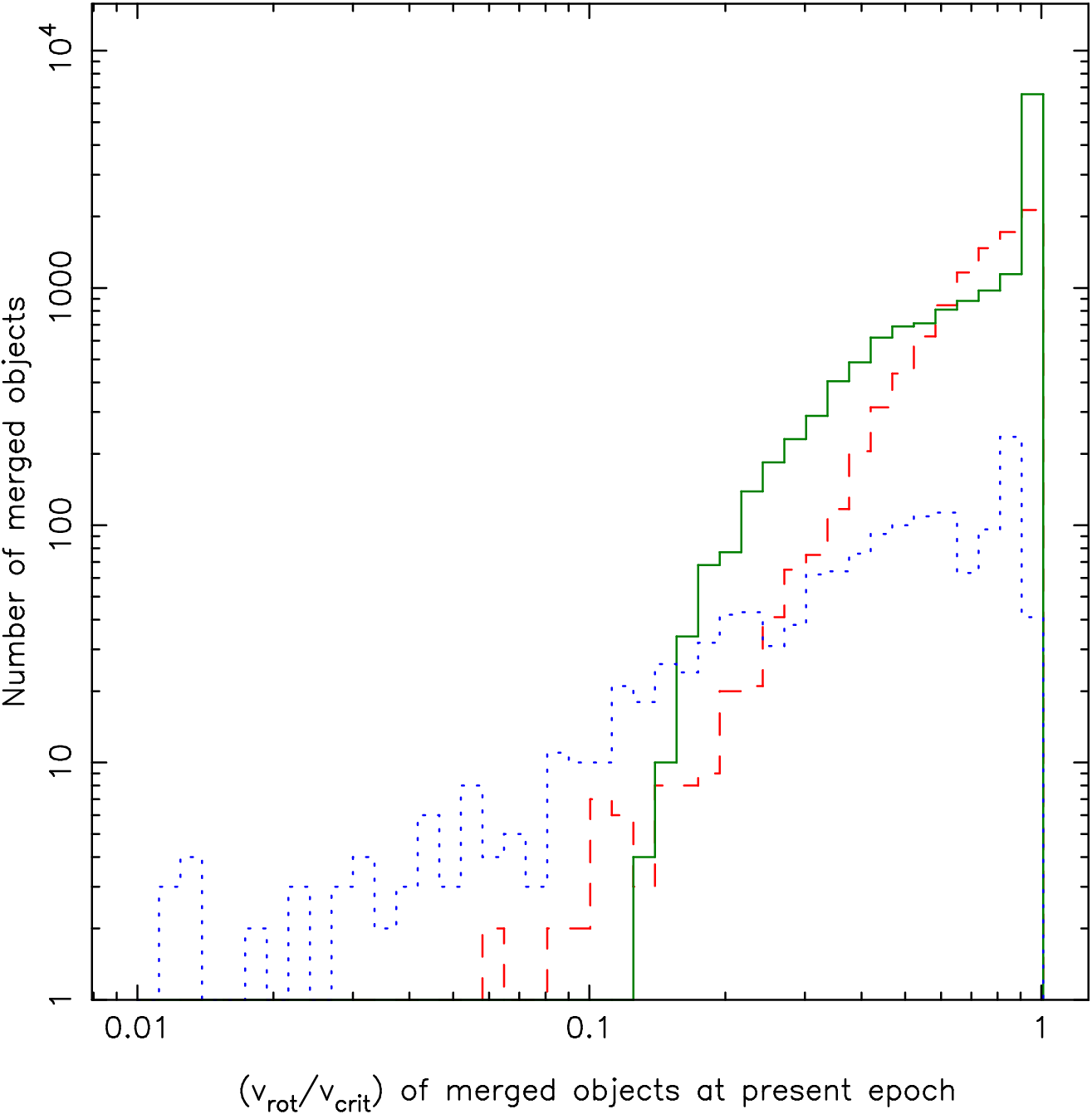}
  \figcaption{
    Distribution of the rotational velocities in our standard model population of present-day merged objects 
    for three sub-populations: RGB stars (dashed curve), HB stars (solid curve) and AGB stars (dotted 
    curve).  The horizontal axis shows the rotational velocity as a fraction of the 
    critical rotational velocity, $v_\mathrm{crit}$, which is assumed to be $\onethird$ of the break-up 
    velocity (see text).
    \label{fig:vrot_1D_merger}
  }
\end{figure}

In Fig.\,\ref{fig:vrot_1D_merger}, we show the predicted distribution of present-day rotational velocities in 
our model population of merged objects for the three sub-populations: RGB stars (red/dashed), HB stars (green/solid) and AGB 
stars (blue/dotted).  The rotational velocity is given as a fraction of an assumed critical 
velocity for rotational mass loss, $v_\mathrm{crit}$ (see \S\,\ref{sec:treatment}). 
For our standard model, $v_\mathrm{crit}$ = 14.6\,km\,s$^{-1} (M/M_\odot)^{1/2} (R/100\Rsun)^{-1/2}$.

We find that the distributions for the AGB, RGB and HB, respectively, are increasingly 
peaked towards high $v_\mathrm{rot}$.  This occurs because most 
of the stars are on the RGB directly after merger 
and are rotating rapidly.  During the evolution on the remainder of the RGB, the rotation of the stars will slow down due 
to their nuclear evolutionary expansion.  When helium 
ignites in the core, the merged objects evolve to the HB, which causes them to shrink and increase their rotational velocities.  Many 
of these stars will reach critical rotation while shrinking, thus forcing them to lose mass and angular momentum as 
described in \S\,\ref{sec:evolution}.  These stars will arrive on the HB spinning critically, causing the peak at 
$\log(v_\mathrm{rot}/v_\mathrm{crit}) \approx 0$.  Since HB stars do not expand dramatically during the core helium burning phase, 
many will be found at or close to this peak.  When these stars continue their evolution on the AGB, they will be spinning 
less rapidly than they were on the RGB at a similar radius, 
because of angular momentum loss during the intervening evolution.
In fact, of the three sub-populations, there exists a relatively sharp lower cutoff in rotational velocities at $0.1\,v_\mathrm{crit}$ 
for both the HB and RGB stars, whereas the AGB stars show a smooth decline. 
For merged objects that are AGB stars at the present epoch, we find that 98\% were formed through 
a merger on the RGB, while only 2\% actually merged on the AGB.  

\begin{figure*}
  \begin{center}
    \includegraphics[width=0.47\textwidth]{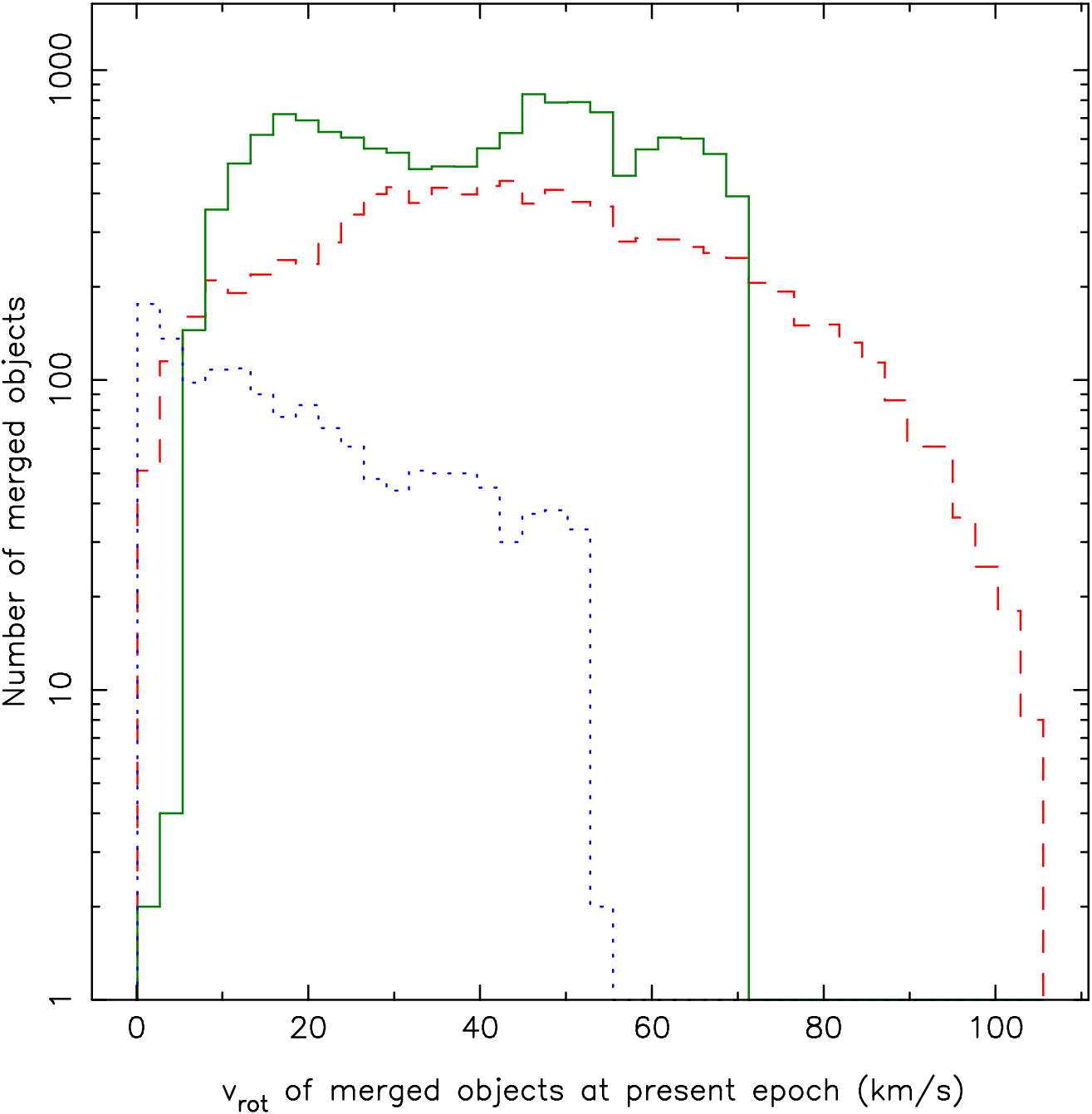}
    \hspace{0.04\textwidth}
    \includegraphics[width=0.47\textwidth]{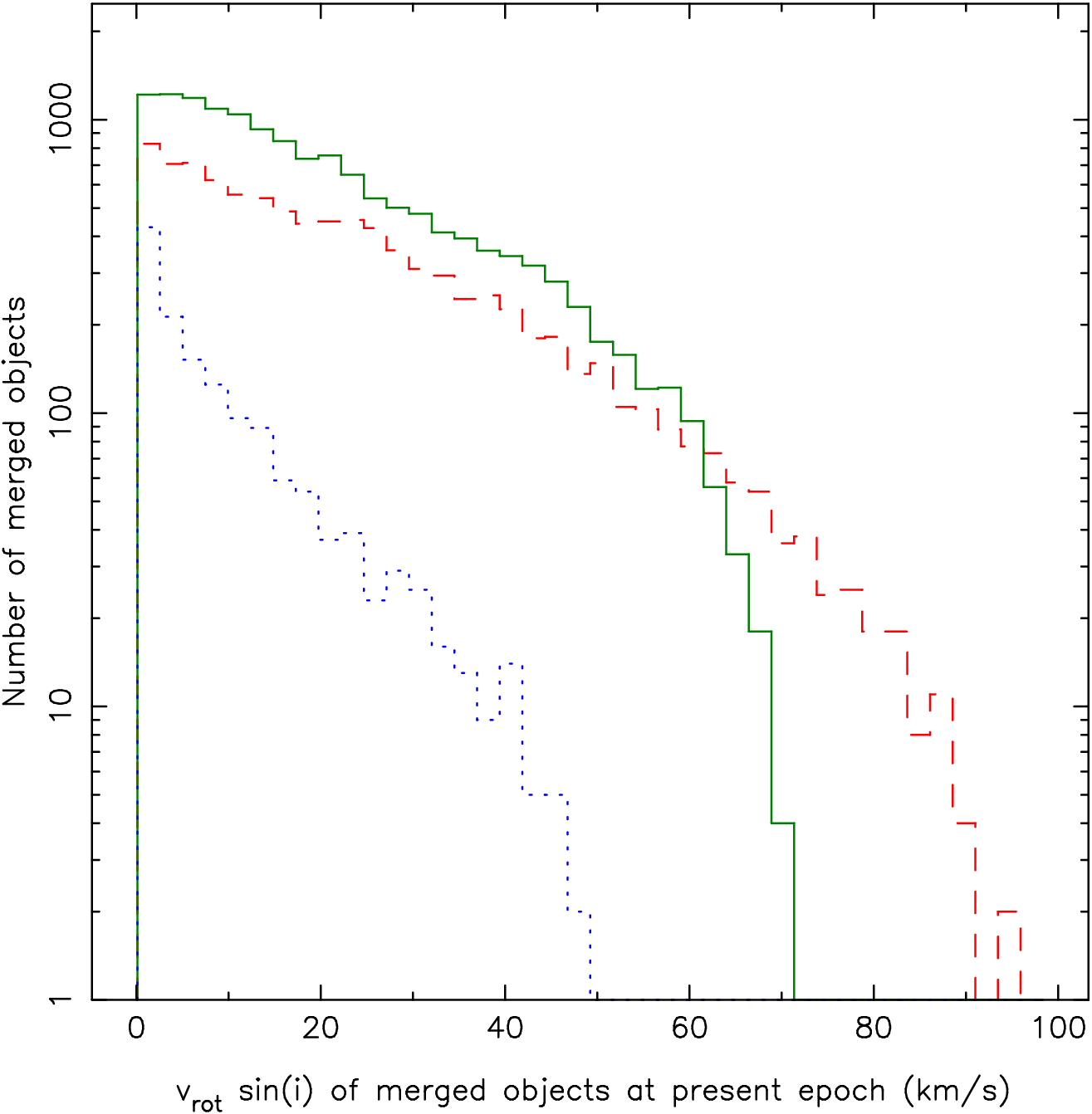}
  \end{center}
  \figcaption{
    Histograms of the equatorial rotational velocities of the sub-populations of merged objects on the 
    RGB (dashed curve), HB (solid curve) and AGB (dotted curve) in our standard model.
    The left-hand panel \textbf{(a)} shows the distributions of \textit{physical} velocities in km/s, 
    the right-hand panel \textbf{(b)} shows the \textit{projected} velocities $v \sin i$ as would be inferred
    from the Doppler shift.
    \label{fig:vrot-kms_1D_merger}
  } 
\end{figure*}

To facilitate comparison with observations,  
Figure~\ref{fig:vrot-kms_1D_merger}a shows the same distributions as Fig.\,\ref{fig:vrot_1D_merger}, but with the velocity 
converted to a physical equatorial velocity in km\,s$^{-1}$.  In addition, in Fig.\,\ref{fig:vrot-kms_1D_merger}b we have computed the
projected rotational velocity, $v \sin i$, as could be inferred from spectroscopic observations.
We used a uniform distribution in $\sin i$ to generate a random inclination for each of our merger remnants.
Whereas most late-type giant stars are observed to rotate with projected velocities of a few km\,s$^{-1}$ \citep[\textit{e.g.},][]{car09}
--- indeed our comparison model of a single-star population (see \S\,\ref{sec:single})
has a median $v \sin i$ of 2.3\,km\,s$^{-1}$ --- we find that the median value for $v \sin i$ of the total merged population 
in Fig.\,\ref{fig:vrot-kms_1D_merger}b is 16.2\,km\,s$^{-1}$.

\begin{figure}
  \plotone{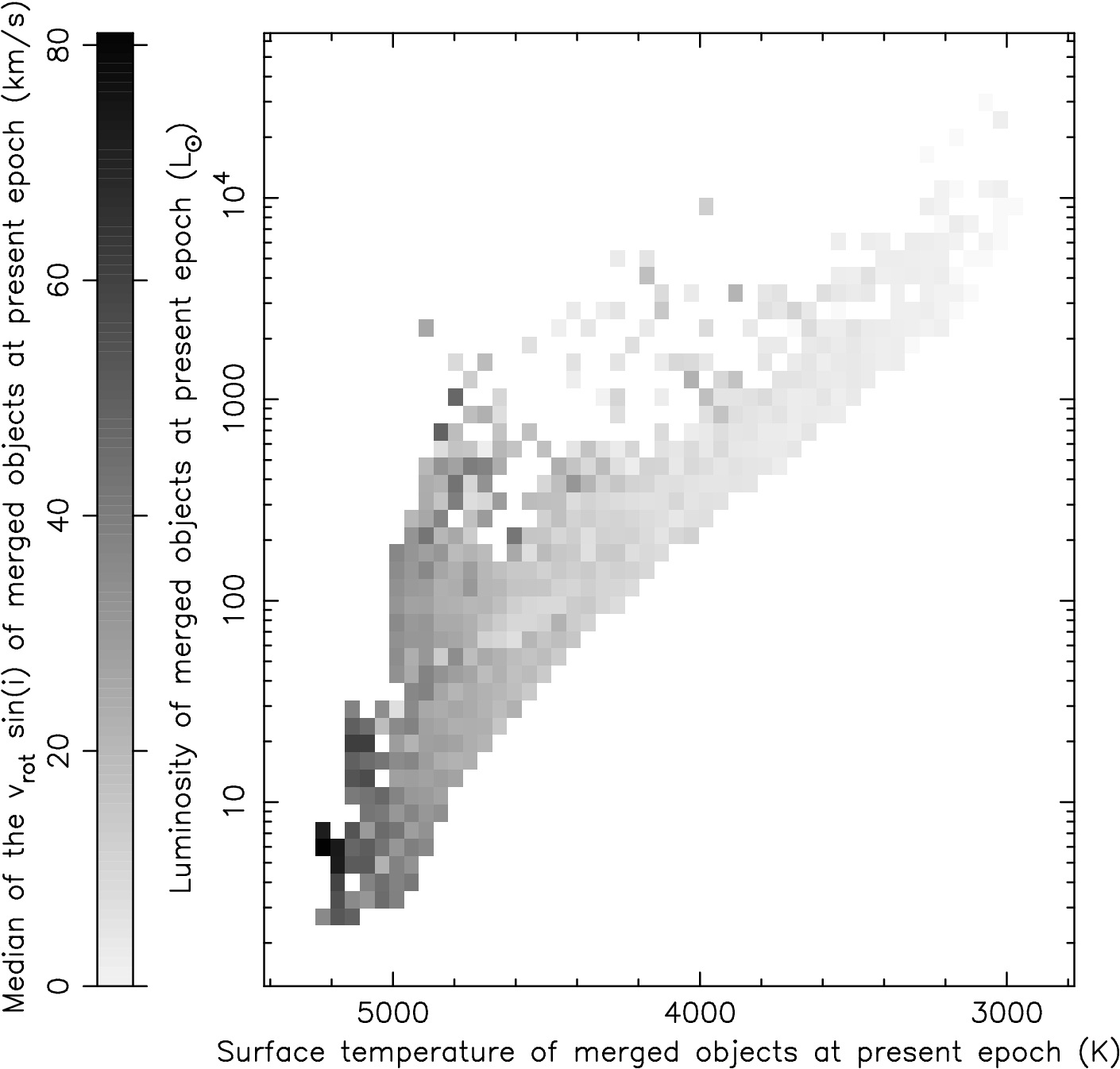}
  \figcaption{
    Two-dimensional histogram of the projected equatorial rotational velocities, $v \sin i$ (in km/s), 
    of the merged objects in our standard model, overlaid on an HRD.
    \label{fig:vrot-kms_2D_merger}
  } 
\end{figure}

In Figure~\ref{fig:vrot-kms_2D_merger}, we show an HR diagram with the distribution of the medians of the 
projected rotational velocities in each bin.  The highest velocity stars are found at the bottom of the RGB,
with physical and projected velocities in excess of 100\,km\,s$^{-1}$ and 80\,km\,s$^{-1}$, respectively.  Stars on the RGB around
5000\,K and on the HB still have projected velocities between 10 and 60\,km\,s$^{-1}$, whereas the merger products in the upper
regions of the HR diagram have rotational velocities that are more common for giants (a few km\,s$^{-1}$), although still 
faster than for single stars in that region of the HR diagram.

\subsubsection{Total mass} 
\label{sec:mtot}

\begin{figure}
  \plotone{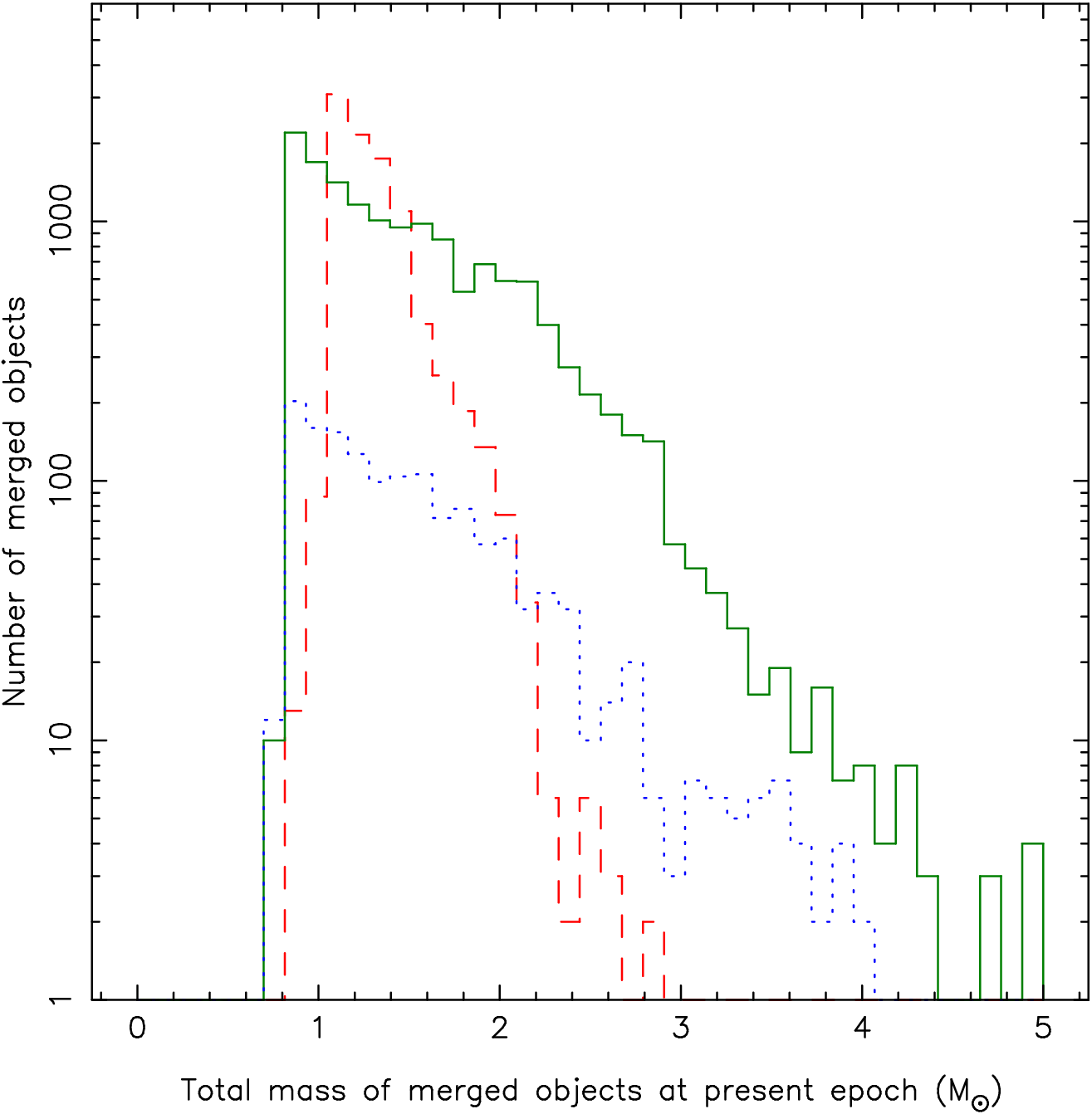}
  \figcaption{
    Distribution of the total masses in our standard model population of present-day merged objects for three 
    sub-populations: RGB stars (dashed curve), HB stars (solid curve) and AGB stars (dotted curve).
    \label{fig:mtot_1D_merger}
  } 
\end{figure}

In Figure~\ref{fig:mtot_1D_merger}, we show the predicted distribution of present-day total masses in our model 
population of merged objects for the three sub-populations:  RGB stars (red/dashed), HB stars (green/solid) and AGB 
stars (blue/dotted).  All three distributions peak at low masses and monotonically decrease towards higher mass,
as may be expected.  However, while the AGB distribution looks much like a scaled-down version of the HB distribution,
the distribution for merged objects presently on the RGB is much narrower, with a higher low-mass cut-off and
a steeper drop-off towards higher masses.  
This steep drop-off occurs because a low-mass star ($M \ltap 2.1\,M_\odot$) with a degenerate 
helium core spends more time on the RGB than a higher-mass star ($M \gtap 2.1\,M_\odot$) with a non-degenerate helium 
core.
As a consequence, of all mergers that occurred while the primary was on the RGB, 42\% of them remain on the RGB 
at the present epoch if $M\ltap 2.1\,M_\odot$, whereas only 1.2\% of them are presently RGB stars if $M \gtap 2.1\,M_\odot$. 
As a result of this, mergers in the latter category end up as rapidly rotating HB stars about 2.5 times more frequently 
and have an 80\% higher chance to be observed as AGB stars today than mergers in the former group.
Table~\ref{tab:dependence} 
shows that, for our standard model, the median mass of the total population is $1.28\,M_\odot$,
with a clearly lower number for the RGB sub-population due to this steep drop-off.

\begin{figure}
  \plotone{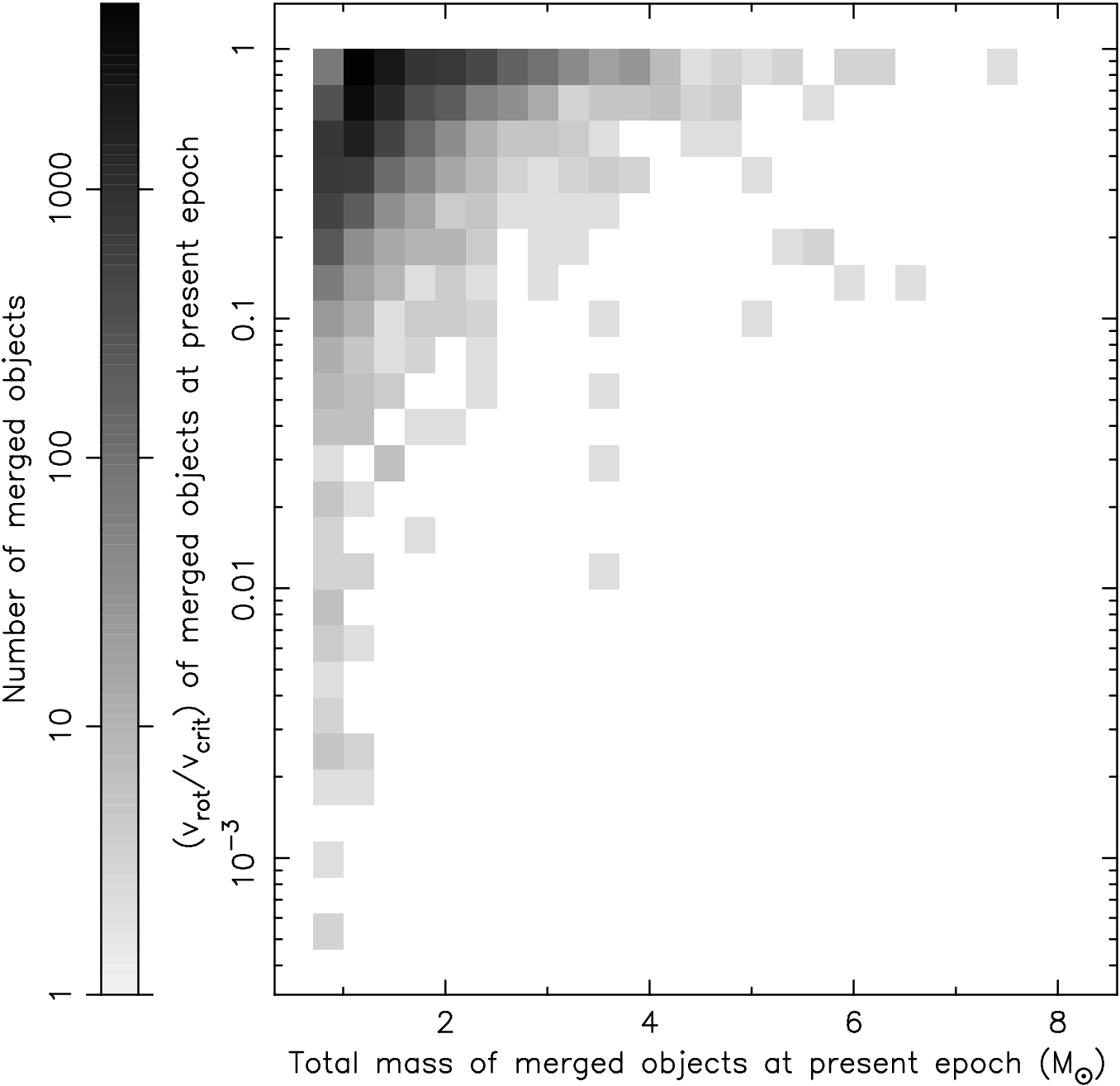}
  \figcaption{
    Two-dimensional distribution of the rotational velocities and total masses in our standard model population 
    of present-day merged objects.  The grey scale represents the total number of merged objects per 
    two-dimensional bin.  As in Fig.\,\ref{fig:vrot_1D_merger}, the rotational velocities are given 
    as a fraction of the assumed critical rotational velocity.
    \label{fig:mtot-vrot_2D_merger}
  } 
\end{figure}

Figure~\ref{fig:mtot-vrot_2D_merger} displays a two-dimensional histogram of the $M_\mathrm{tot}$--$v_\mathrm{rot}$
plane, for the total present-day population of merged objects (RGB, HB and AGB).  This figure indicates that
most of the merger remnants are low-mass ($\ltap 2\Msun$), rapidly rotating stars, with a peak in the distribution 
corresponding to a mass and velocity range of $M \approx 1.0-1.3\,M_\odot$ and $v_\mathrm{rot}/v_\mathrm{crit} \approx
0.75 - 1$, respectively. 

\subsubsection{Luminosity} 
\label{sec:lum}

\begin{figure}
  \plotone{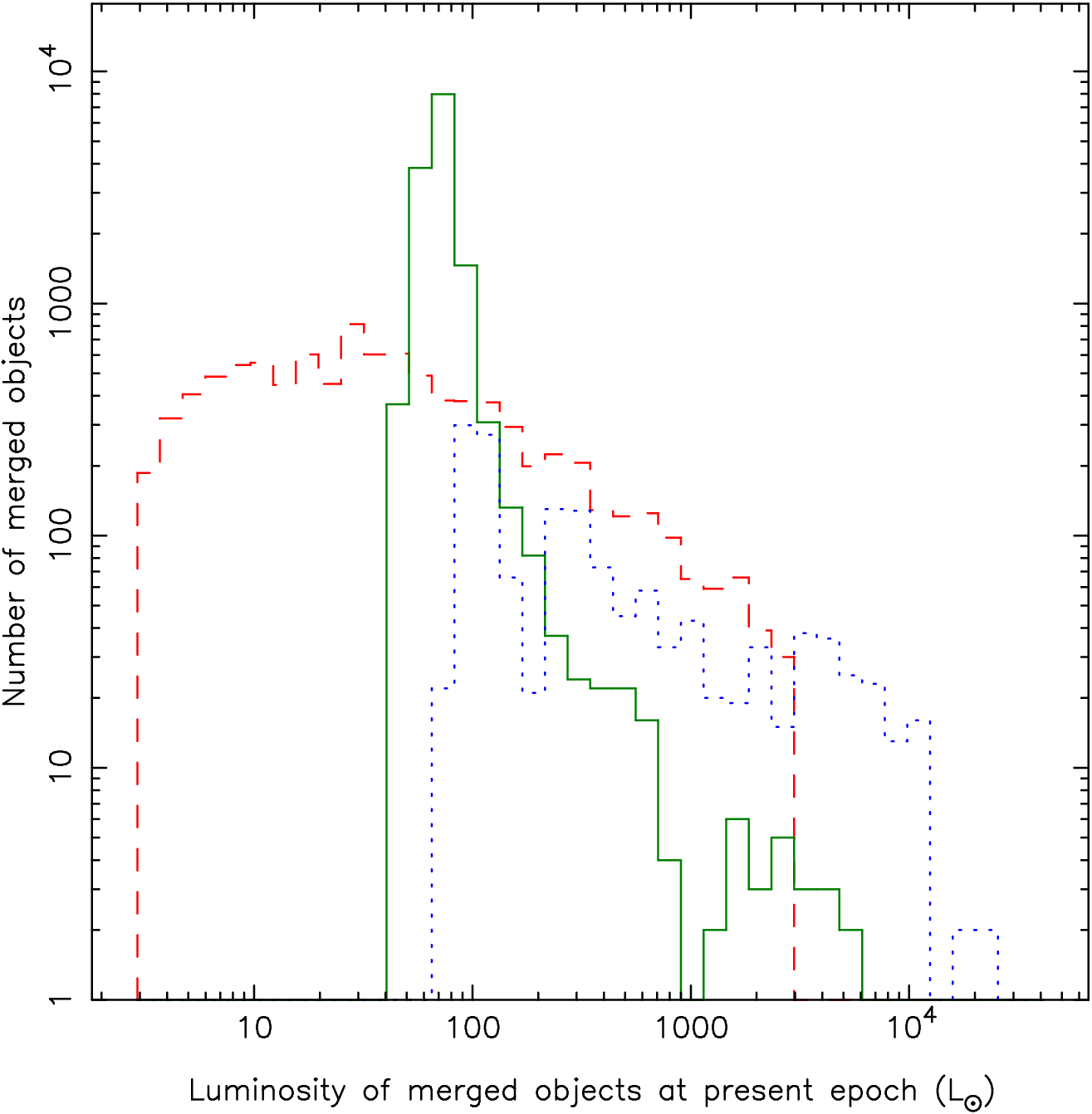}
  \figcaption{
    Distribution of the luminosities in our standard model population of present-day merged objects for three 
    sub-populations: RGB stars (dashed curve), HB stars (solid curve) and AGB stars (dotted curve).
    \label{fig:lum_1D_merger}
  } 
\end{figure}

In Fig.\,\ref{fig:lum_1D_merger}, the predicted distribution of the present-day luminosities in our model population of 
merged objects is illustrated.  As in the previous figures, the distribution is divided into 3 sub-populations: 
RGB stars (red/dashed), HB stars (green/solid) and AGB stars (blue/dotted).  The HB star distribution ranges in 
$\log L/L_\odot$ from 1.6 to $\sim 3$ and is sharply peaked near $\log L/L_\odot = 1.8$. 
The RGB distribution is broader and 
flatter, ranging in $\log L/L_\odot$ from $\sim 0.4$ to 3.3 and having a broad maximum near $\log L/L_\odot = 1.4$.  The AGB 
distribution ranges in $\log L/L_\odot$ from $\sim 1.9$ to 4.1 and has multiple maxima, with a strong peak at 
$\log L/L_\odot \sim 2.0$ 
and a weaker one near $\log L/L_\odot \sim 2.4$.  We note that the strong peak in the AGB distribution lies near that 
of the HB distribution, thus contributing to the peak in the overall luminosity distribution. 

\subsubsection{HR diagram} 
\label{sec:hrd}

\begin{figure*}
   \begin{center}
    \includegraphics[width=0.47\textwidth]{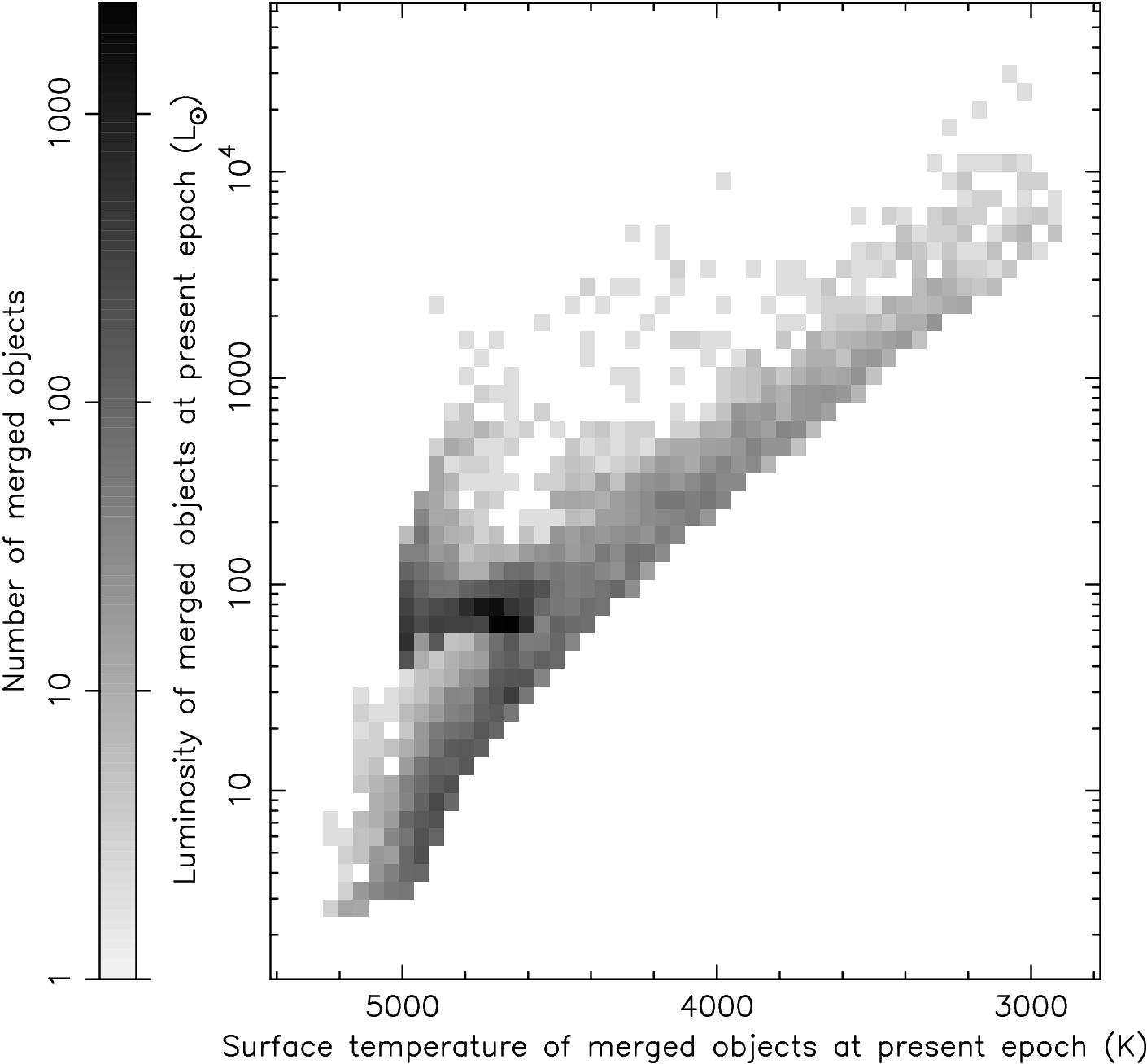}
    \hspace{0.04\textwidth}
    \includegraphics[width=0.47\textwidth]{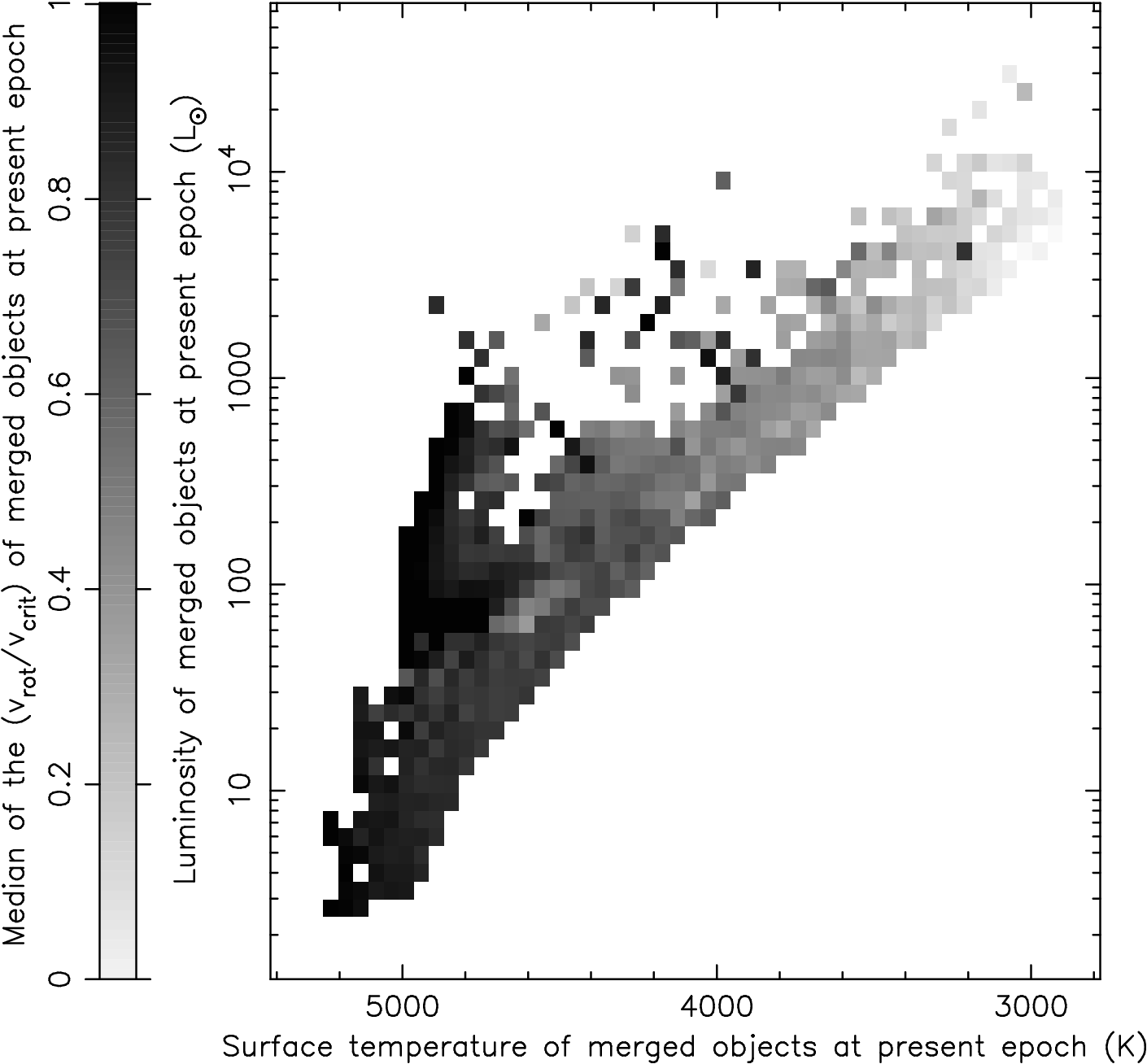}
  \end{center}
  \figcaption{
    Two-dimensional distribution of the luminosities and effective temperatures in our standard model population 
    of present-day merged objects.  In the left-hand panel \textbf{(a)}, the grey scale represents the total number of 
    merged objects per two-dimensional bin (pixel).  The lightest grey scale, which represents a bin with
    one star, is easily distinguished from the white background.  
    In the right-hand panel \textbf{(b)}, the grey scale represents the median of the distribution of rotational velocities 
    (as a fraction of $v_\mathrm{crit}$) per pixel.  Temperature increases to the left on the horizontal 
    axis to facilitate comparison with observational HR diagrams.
    \label{fig:hrd_merger}
  } 
\end{figure*}

In Figure~\ref{fig:hrd_merger}, we show binned HR diagrams (HRDs) for the total model population of merged objects.  
Fig.\,\ref{fig:hrd_merger}a is a two-dimensional histogram, in which the grey scale indicates the number of merged 
objects per two-dimensional bin (pixel).  If we draw an imaginary line from the upper-left to the lower-right 
corners of the diagram, this line intersects the distribution on the giant branches almost perpendicularly.  When 
we start in the upper-left corner and follow this line through the distribution, we find that there is a gradient 
in the number density of merged objects, and that this density increases as we continue to follow the line towards 
the lower-right corner.  This gradient is the result of the IMF; there are more lower-mass than higher-mass stars, 
and the evolutionary tracks of lower-mass stars on the giant branches lie to the lower right of those of higher-mass 
stars.  In addition, we find that the majority of our merged objects sit in a dense clump at the HB (see \S\,\ref{sec:vrot}), 
with $T_\mathrm{eff} \approx 4500-5000$\,K and $L\approx60\,L_\odot$.

Figure~\ref{fig:hrd_merger}b displays the medians of the velocity distributions of merged objects in each bin of 
the HRD.  Since the standard deviations of these distributions are almost always significantly smaller than the
medians, this is a relevant estimate of the rotational velocity.  Whereas the number of stars in each bin is a 
function of mass, the velocity is clearly dependent on evolutionary phase.  Especially for the RGB, it is clear
that the rotational velocity diminishes towards the tip of the giant branch.  The HB contains fast rotators, as 
we saw earlier, and even the AGB shows a few pixels with a high rotational velocity, although Fig.\,\ref{fig:hrd_merger}a
indicates that these pixels each contain about one star.

\subsubsection{Oblateness} 
\label{sec:oblate}

The high rotational velocities of the present-day merged objects (see \S\,\ref{sec:vrot}) may well give 
rise to deformation of the stars.  In order to quantify the oblateness of the stars, we assumed they can be treated
as MacLaurin spheroids, for which the relation between angular frequency $\omega$, density $\rho$ and oblateness 
$e$ is given by
\begin{equation}
  \frac{\omega}{\sqrt{2 \pi G \rho\,}} \!=\! 
  \left[\!\frac{\left(1\!-\!e^2\right)^{\!1/2}}{e^3} \left(3\!-\!2e^2\right) \sin^{\!-1}(e)  - \frac{3}{e^2}\!\left(1\!-\!e^2\right)\!\right]^{\!1/2}\!,
  \label{eq:maclaurin}
\end{equation}
\citep{maclaurin}, where we use $\rho = \bar{\rho}_*$, the mean density of the star, and the eccentricity or 
oblateness is defined in terms of the polar and equatorial radius ($r_\mathrm{p}$ and $r_\mathrm{e}$, 
respectively) as
\begin{equation}
  e = \sqrt{1-\left(\frac{r_\mathrm{p}}{r_\mathrm{e}}\right)^2}.
  \label{eq:defe}
\end{equation}

We note that for a constant density, this function resembles a straight line for the regime of our interest 
($e<0.85$) and we use a simple least-squares method to find that the best slope for a line through the origin 
($e=0, \omega=0$) to describe this part of the function (to an accuracy of 2.5\% or better) is 0.53.  This 
allows us to derive an approximation for the oblateness as a function of the angular frequency and the density
\begin{equation}
  e \approx \frac{\omega}{0.53 \, \sqrt{2 \pi G \rho\,}} = \frac{0.75 \, \omega}{\sqrt{G \rho\,}}; ~~~ e<0.85
  \label{eq:oblateness}
\end{equation}
and to compute the oblateness for each of the merger products in our standard population.

\begin{figure*}
  \begin{center}
    \includegraphics[width=0.43\textwidth]{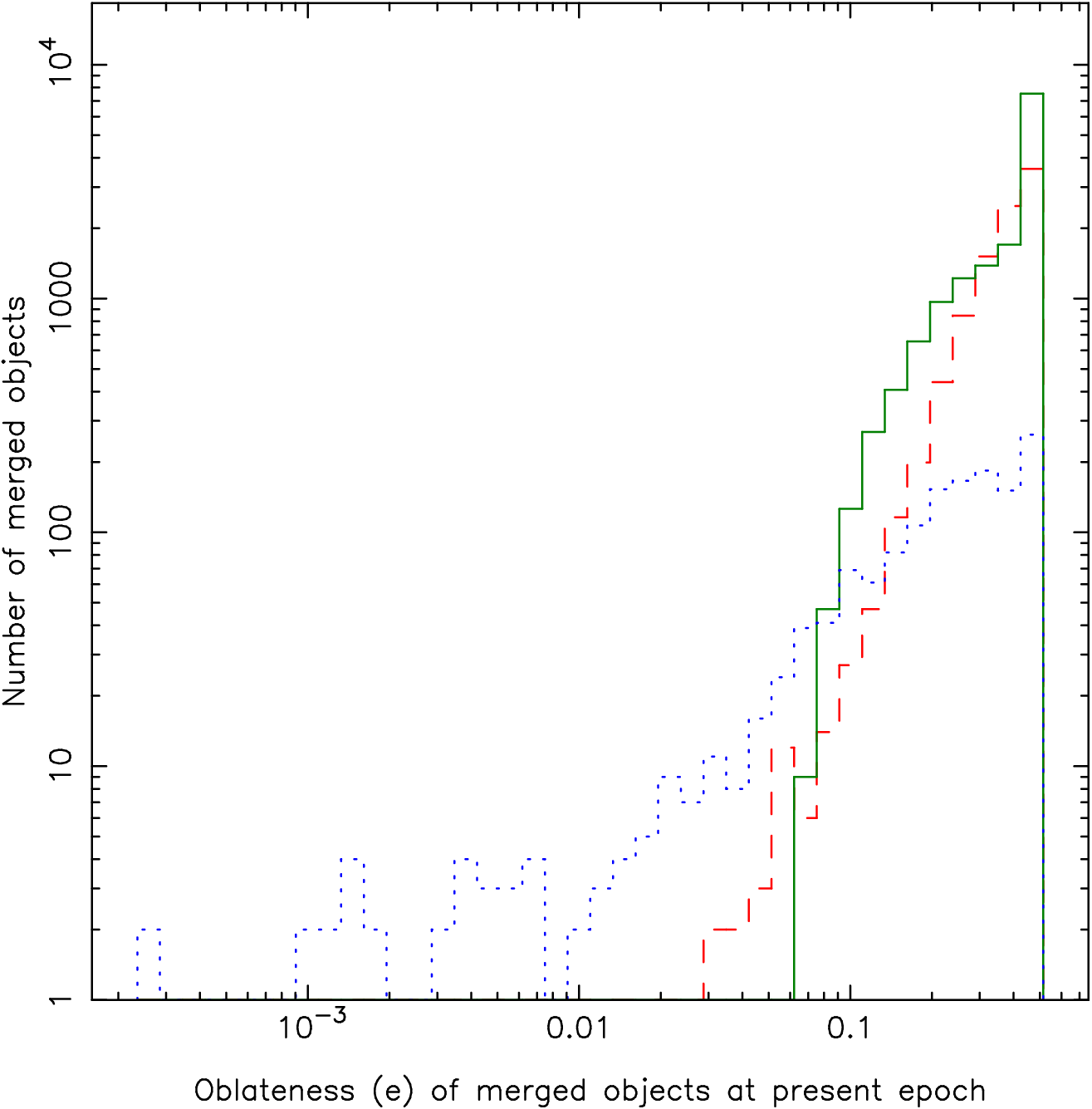}
    \hspace{0.055\textwidth}
    \includegraphics[width=0.47\textwidth]{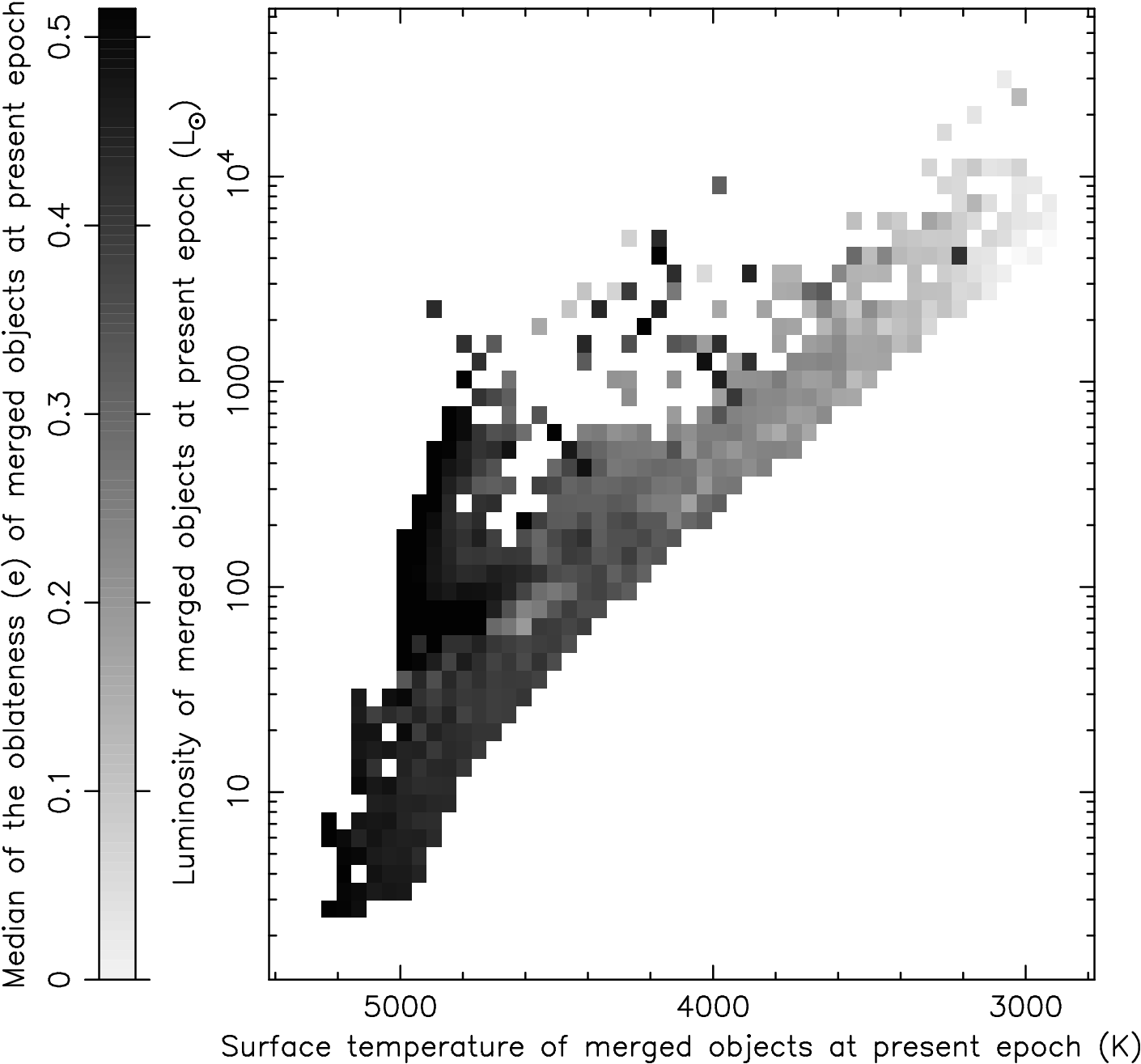}
  \end{center}
  \figcaption{
    Distributions of the oblateness $e$ (see Eq.\,\ref{eq:defe}) in our model population of present-day merged 
    objects. 
    Left-hand panel \textbf{(a)}: histograms for the sub-populations of RGB stars (dashed line), HB stars (solid line) and AGB stars 
    (dotted line).
    Right-hand panel \textbf{(b)}: two-dimensional distribution of the oblateness on an HRD.  The stars at the base of the 
    RGB are most deformed, whereas very large giants are nearly spherical.
    \label{fig:oblateness}
  } 
\end{figure*}

Figure~\ref{fig:oblateness}a shows the distribution of the oblateness of the stars in our present-day
populations of RGB, HB and AGB stars.  We find that for most stars the eccentricity is larger than $0.1$ and
should be observable if the star is not too distant \citep[see \textit{e.g.}][]{zhao09}. All three distributions 
peak at the maximum value of the oblateness ($e \approx 0.5$).  
While the histograms for the RGB and HB drop off rapidly towards smaller oblateness, the
histogram for AGB stars does so less steeply and there are a small number of AGB stars with a small eccentricity
($e<0.01$).
Figure~\ref{fig:oblateness}b displays the distribution of eccentricities on an HRD and
shows that the most deformed stars are at the base of the red-giant branch, the HB and the lower AGB,
while the most expanded giants have shapes that are closer to spherical.  The distributions of eccentricities
follow those of $\log(v_\mathrm{rot}/v_\mathrm{crit})$ in Figs.\,\ref{fig:vrot_1D_merger} and 
\ref{fig:hrd_merger}b. 
We stress that our model of a MacLaurin spheroid for the shape of these stars has
limited validity and the fit in Eq.\,\ref{eq:oblateness} depends on our choice of $\beta$. For example, if
$\beta=1$, this fit, and the assumption of MacLaurin spheroids, would yield physically meaningless results.
Consequently, our use of MacLaurin spheroids is meant to provide rough order-of-magnitude estimates and trends,
not precise values.

\subsection{Dependence upon input parameters}
\label{sec:dependence}

In addition to our standard model, we investigated the dependence of our results on three 
uncertain input parameters:  the CE efficiency parameter, $\alpha_\mathrm{CE}$, 
the initial mass ratio distribution in ZAMS binaries, $g(q)$, and the assumed critical 
rotational velocity at which mass is lost, $v_\mathrm{crit}$ (see item 1 of \S\,\ref{sec:treatment}).
Uncertainties in the first two 
parameters plague the majority of population synthesis calculations involving close binary systems, while uncertainty 
in the third parameter is specific to the present calculations. 
We calculated model populations for two other choices of each parameter and have
listed the parameter choices for each model in Table~\ref{tab:models}.
In all cases, only one parameter was varied at a time, keeping the other two at their default values.

Selected results from these non-standard models for the three sub-populations discussed in \S\,\ref{sec:properties} and for the total population
are shown in Table~\ref{tab:dependence}.  We find that, with a few notable exceptions, varying these input parameters has relatively little effect on the overall characteristics of the present-day 
population of merged objects.  For example, the fraction of the population that is rotating slowly ($v_\mathrm{rot} \leq 0.1\,v_\mathrm{crit}$) 
is uniformly small for all choices of input parameters:  in the RGB and HB sub-populations, this fraction is either zero or so small as not to be statistically 
significant; in the AGB sub-population, this fraction is 7 -- 9\%; and for the total population, this fraction is barely significant ($<1$\%). Further, with the 
exception of the $v_\mathrm{crit} = v_\mathrm{br}$ model, the 
present-day median mass in the total population varies from $1.20\Msun$ to $1.29\Msun$ and the variation within a given sub-population is only slightly greater.  The 
median mass in the $v_\mathrm{crit} = v_\mathrm{br}$ model is systematically greater than the others because 
the merged object can assimilate more angular momentum (and hence mass) during the merger before it reaches its critical rotation velocity 
(see assumption 2 in \S\,\ref{sec:treatment}).  
As expected, the median projected rotational velocity, $v \sin i$, is strongly correlated with our assumed choice for $v_\mathrm{crit}$;
a larger value for $v_\mathrm{crit}$ allows a higher median projected rotational velocity.
There is somewhat greater variation among the models in the fraction of present-day merged objects rotating at $v_\mathrm{crit}$.  In the total population, this 
fraction varies by a factor of 2, from 0.13 for the $v_\mathrm{crit} = v_\mathrm{br}$ model to 0.27 for the $\alpha_\mathrm{CE} = 0.1$ model, compared with 
0.19 for our standard model.  The same trend is found in the RGB and HB sub-populations. 
Virtually none of the AGB stars are rotating critically, because of the large rotational inertia of the envelopes in these stars.

The largest variation within the present-day population caused by changing the input parameters is found in the 
total number of merged objects.  For the total population and within a given sub-population, the total number of present-day merged objects varies 
by roughly 20\% to 50\%, with the smallest number of objects found in the $v_\mathrm{crit} = v_\mathrm{br}$ model population 
(24,414 stars, or 0.24\% of the input population) and the largest (32,882 stars, or 0.33\% of the input population) in the $\alpha_\mathrm{CE} = 0.1$ model
(except for RGB stars, where the 
largest number is found in the $v_\mathrm{crit} = 0.1\,v_\mathrm{br}$ model).  Reducing the value of $\alpha_\mathrm{CE}$ results in a less efficient transfer of 
orbital energy to the CE during spiral-in, thereby increasing the number of systems that merge during the CE phase.  This increase in the number of CE 
mergers results in a larger number of present-day merged objects.  
As one might expect, varying  $v_\mathrm{crit}$ has very little effect on the total number of present-day merged objects. 

Column 8 of Table~\ref{tab:dependence} lists the mass that is not assimilated from the binary during merger, $M_\mathrm{rej}$, and column 9 lists 
this mass as a fraction of the 
total mass of the binary at the beginning of the CE phase. We find that varying the choice of $\alpha_\mathrm{CE}$ has essentially no effect on the amount of 
mass rejected during merger.  Increasing $v_\mathrm{crit}$ decreases $M_\mathrm{rej}$, which is as expected since 
for larger values of  $v_\mathrm{crit}$, 
the merged object can assimilate a greater fraction of the mass of the binary before it starts to spin critically.
Perhaps surprisingly at first, varying $g(q)$ has the strongest effect on the value of $M_\mathrm{rej}$.
However, this is simply because $g(q)$ determines the 
ZAMS distribution of secondary masses.  For $g(q) = q$, equal mass ZAMS binaries are strongly favored, resulting in systematically larger secondary 
masses than in our standard model.  Conversely, for $g(q) = q^{-0.9}$, extreme mass ratio ZAMS binaries are strongly favored, resulting in 
systematically smaller secondary masses than in our standard model.  
Consequently, for $g(q) = q$, $M_\mathrm{rej}$ is larger than in our standard model because the secondaries are more massive.  
On the other hand, for $g(q) = q^{-0.9}$, $M_\mathrm{rej}$ is smaller than in our standard model because the secondaries are less massive.

The last column in Table~\ref{tab:dependence} lists the fraction of mass lost by the merged objects between the moment immediately after 
merger and the present
epoch. This fraction is vanishingly small for present-day RGB stars, but is $\sim$12\% for objects that are currently on the HB and AGB, 
indicating that this post-merger mass loss occurs predominantly 
near the moment of core helium ignition.  
Two factors contribute to this mass loss: (1) normal stellar winds near the tip of the RGB \textit{prior} to helium ignition and (2) rotationally-enhanced mass loss during contraction 
to the HB \textit{after} helium ignition.
A typical RGB star spends most of its time on the lower part of the RGB without much mass loss.  Then, in a relatively short time,
the star evolves to the tip of the RGB, where mass loss becomes significant.  For example, in the 1.3$\Msun$ stellar model in our grid, $\sim0.2\Msun$ is lost on the RGB, with 90\% of this mass lost in the last 4\% of the time spent on the RGB.  While wind mass loss on the RGB is present whether or not merger occurs, we find that for a substantial number of objects that merge on the RGB, additional mass loss occurs during the star's contraction after helium ignition.
Especially in stars that undergo a helium flash, where the decrease in stellar radius is large, the rotational velocity of the merged object 
becomes critical as the star contracts to the HB.  A period of enhanced mass loss ensues,
until the star has lost a sufficient amount of angular momentum
for the rotational velocity to drop below critical.  
While we do not distinguish between these two contributions 
in column 10 of Table~\ref{tab:dependence},
we note that the post-merger mass loss is  
lower in the $v_\mathrm{crit} = v_\mathrm{br}$ model since the larger value for 
$v_\mathrm{crit}$ allows the stars to reach larger rotational velocities before they start to experience
rotationally-enhanced mass loss during contraction to the HB.

\subsection{Comparison with single stars}
\label{sec:single}

\begin{figure}
  \plotone{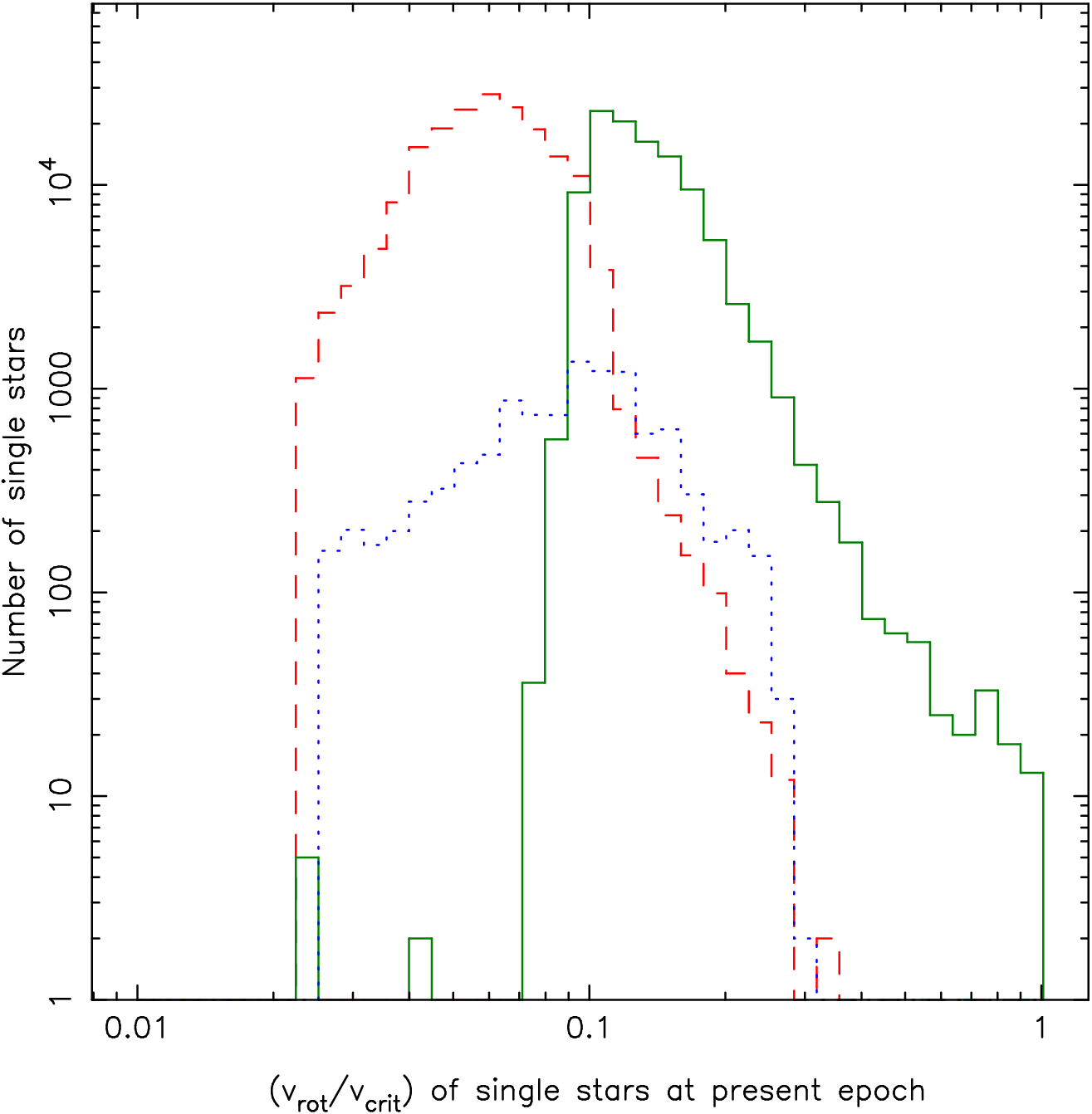}
  \figcaption{
    Distribution of the rotational velocities in a comparison model of normal present-day single stars 
    constructed from the same grid of stellar evolution models used for the merged population. As in the 
    corresponding plot for the merged population (Fig.\,\ref{fig:vrot_1D_merger}), the rotational velocities 
    are given as a fraction of the assumed critical rotational velocity and the distribution is separated into 
    three sub-populations: RGB stars (dashed curve), HB stars (solid curve) and AGB stars (dotted curve).
    \label{fig:vrot_1D_single}
  }
\end{figure}

In this section, we compare our present-day model population of merged objects to a model population of normal present-day single 
stars.  We generated $10^7$ single ZAMS stars, drawing their initial masses from the same IMF that was used for the merged objects, 
and determined the structure and evolution of these stars using the same grid of stellar models that was used to generate the merged object 
population (see \S\,\ref{sec:code}).  We consider only those present-day single stars that are on the RGB, HB or AGB for this comparison.
In Fig.\,\ref{fig:vrot_1D_single}, we show the present-day distribution of rotational velocities for our single stars in each of 
these evolutionary phases.
As before, the rotational velocity for a given star is shown as a fraction of the critical rotational velocity for that star, defined as in our 
standard model
($v_\mathrm{crit} = \onethird\,v_\mathrm{br}$).  The present-day rotational velocities for single stars were calculated using the following expression 
for rotational velocities in ZAMS stars as a function of mass \citep[][Eq.\,107]{hur00},
\begin{equation}
v_\mathrm{rot,ZAMS} = \frac{330\,M^{3.3}}{15.0+M^{3.45}}~\mathrm{km\,sec}^{-1},
\label{eq:rotZAMS}
\end{equation} conserving angular momentum up to the base of the RGB, and then accounting for angular momentum 
loss from the star via stellar winds throughout the remainder of the evolution.  Once again, angular momentum loss due to magnetic braking was not 
incorporated in the evolution. 

\begin{figure}
  \plotone{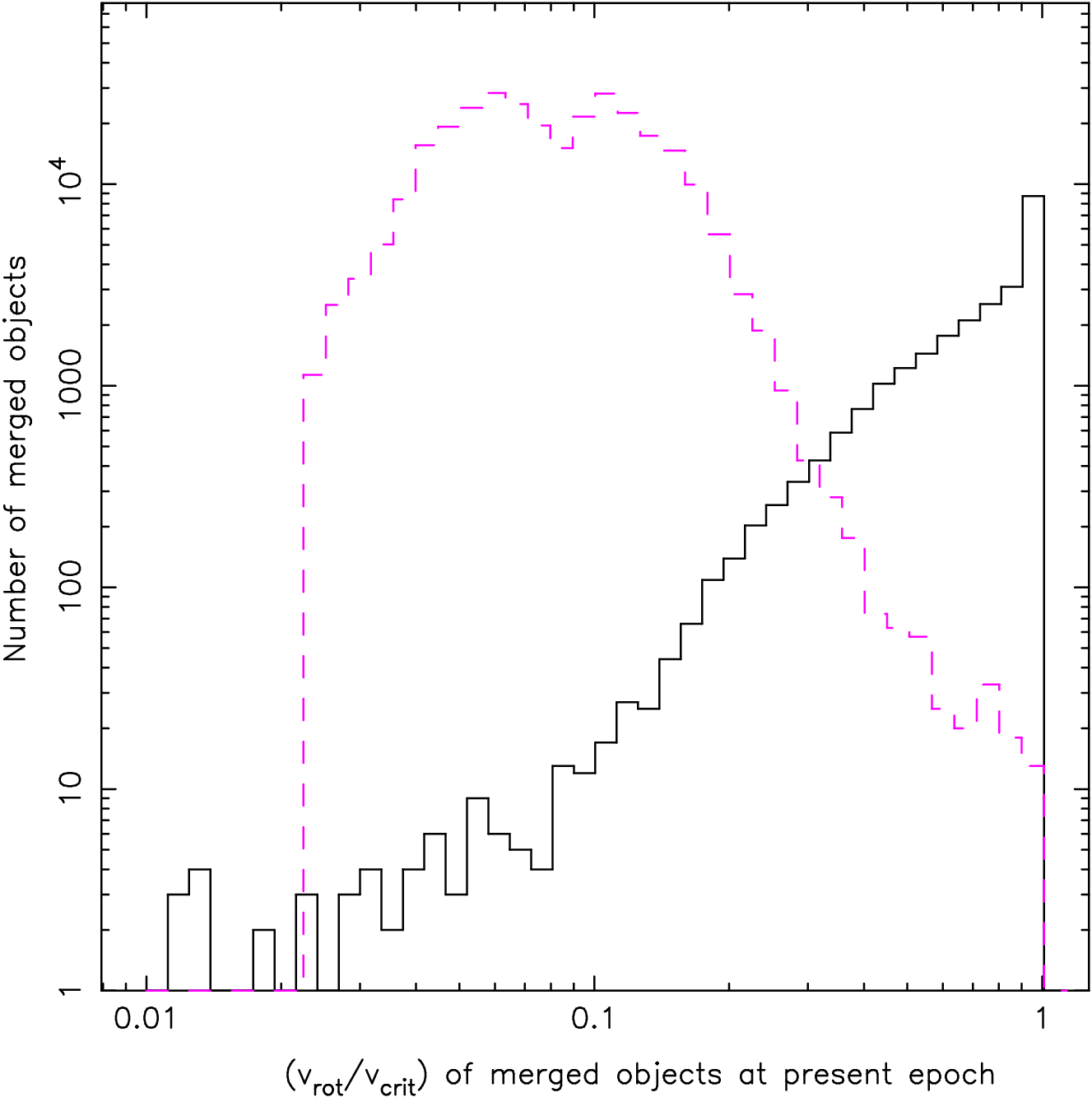}
  \figcaption{
    Distribution of the rotational velocities in our model population of present-day merged objects (solid curve) 
    and in a corresponding present-day population of normal single stars (dashed curve) calculated using the same 
    grid of stellar evolution models.  The solid curve represents the sum of the combined sub-populations of 
    RGB, HB and AGB stars shown in Fig.\,\ref{fig:vrot_1D_merger} and the dashed curve represents the sum of the 
    combined sub-populations of RGB, HB and AGB stars shown in Fig.\,\ref{fig:vrot_1D_single}.  As in those two figures, the rotational 
    velocities here are given as a fraction of the assumed critical rotational velocity.
    \label{fig:vrot_1D_compare}
  } 
\end{figure}

Comparison of Fig.\,\ref{fig:vrot_1D_single} with the corresponding plot for the merged objects (Fig.\,\ref{fig:vrot_1D_merger}) 
reveals a striking contrast in the rotational velocities for the two 
populations.  The peak rotational velocity for each sub-population of merged objects is approximately an order of magnitude higher than the 
corresponding peak rotational velocity for normal single stars.  This difference can be seen even more clearly in Fig.\,\ref{fig:vrot_1D_compare}, 
which shows the distribution of rotational velocities for the total populations of merged objects and single stars.  
Not only are the peak rotational velocities markedly different, but the overlap between the two distributions 
is relatively small.  Consequently, observational determinations of rotational velocities in RGB, HB and AGB stars may be expected to provide an 
important diagnostic in determining which stars are the result of a CE merger.

\begin{figure}
  \plotone{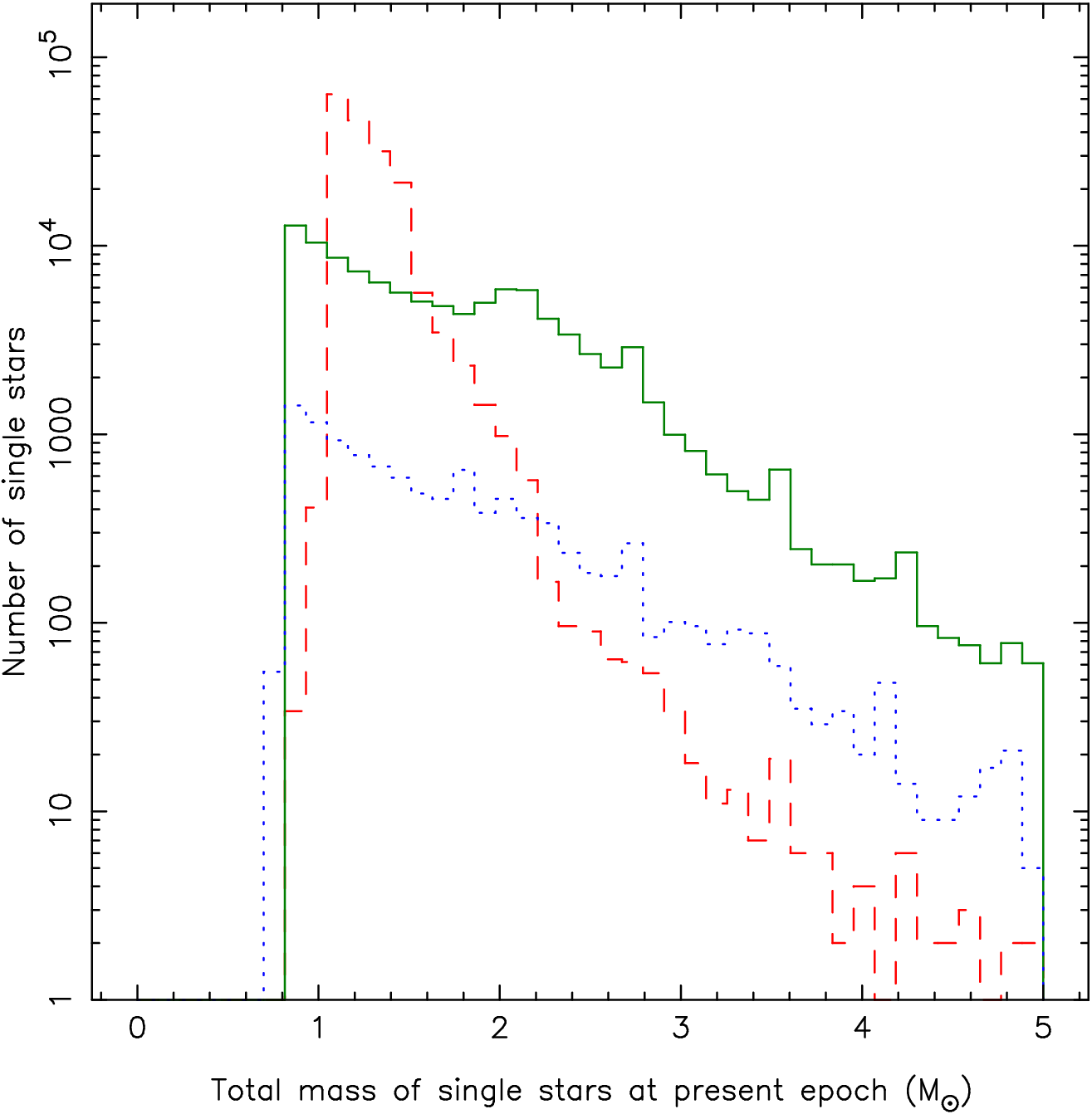}
  \figcaption{
    Distribution of the total masses in a comparison model of normal present-day single stars constructed 
    from the same grid of stellar evolution models used for the merged population. 
    As in the corresponding plot for the merged population (Fig.\,\ref{fig:mtot_1D_merger}), 
    the distribution is separated into three sub-populations: RGB stars (dashed curve), HB stars (solid curve) and AGB stars (dotted curve).
    \label{fig:mtot_1D_single}
  } 
\end{figure}

In Fig.\,\ref{fig:mtot_1D_single}, we show the present-day distribution of masses for RGB, HB and AGB stars in our single star population.  Comparison 
of this plot with the corresponding plot for the merged objects (Fig.\,\ref{fig:mtot_1D_merger}) reveals that 
the merged objects have a slightly steeper drop-off towards higher masses than the single stars, which is due to rotationally-enhanced 
mass loss from many of the merger products during their post-merger evolution. 
However, as can be seen by comparing the median present-day masses listed in Table~\ref{tab:dependence} for 
the total population in our standard model and the single-star population, this difference is small ($0.05\Msun$).  
On the other hand, the relative percentages of RGB, HB and AGB stars are quite different for the two populations.   While the 
merged object population consists of 37\% RGB stars, 57\% HB stars and 5.7\% AGB stars for our standard model, the respective percentages for the
normal single star population are 61\%, 36\% and 3.6\%.  The ratios of these percentages are 0.61, 1.6 and 1.6, respectively,
which suggests that this is caused by a shift in numbers between the pre- and post-helium-ignition phases.
In other words, in a present-day population of merged objects on the
RGB, HB and AGB, the majority of stars have ignited helium, while in a
corresponding normal single-star population consisting of the same three sub-populations, the majority of stars have not yet
ignited helium.
This is understandable since
for the merged objects, we only consider a star \textit{after} it has merged, which is in 99.9\% of the cases 
somewhere on the RGB (see \S\,\ref{sec:numbers}).  Hence, its subsequent evolution as a merged object consists of only 
\textit{part} of the RGB, but the \textit{full} HB and AGB.

\begin{figure}
  \plotone{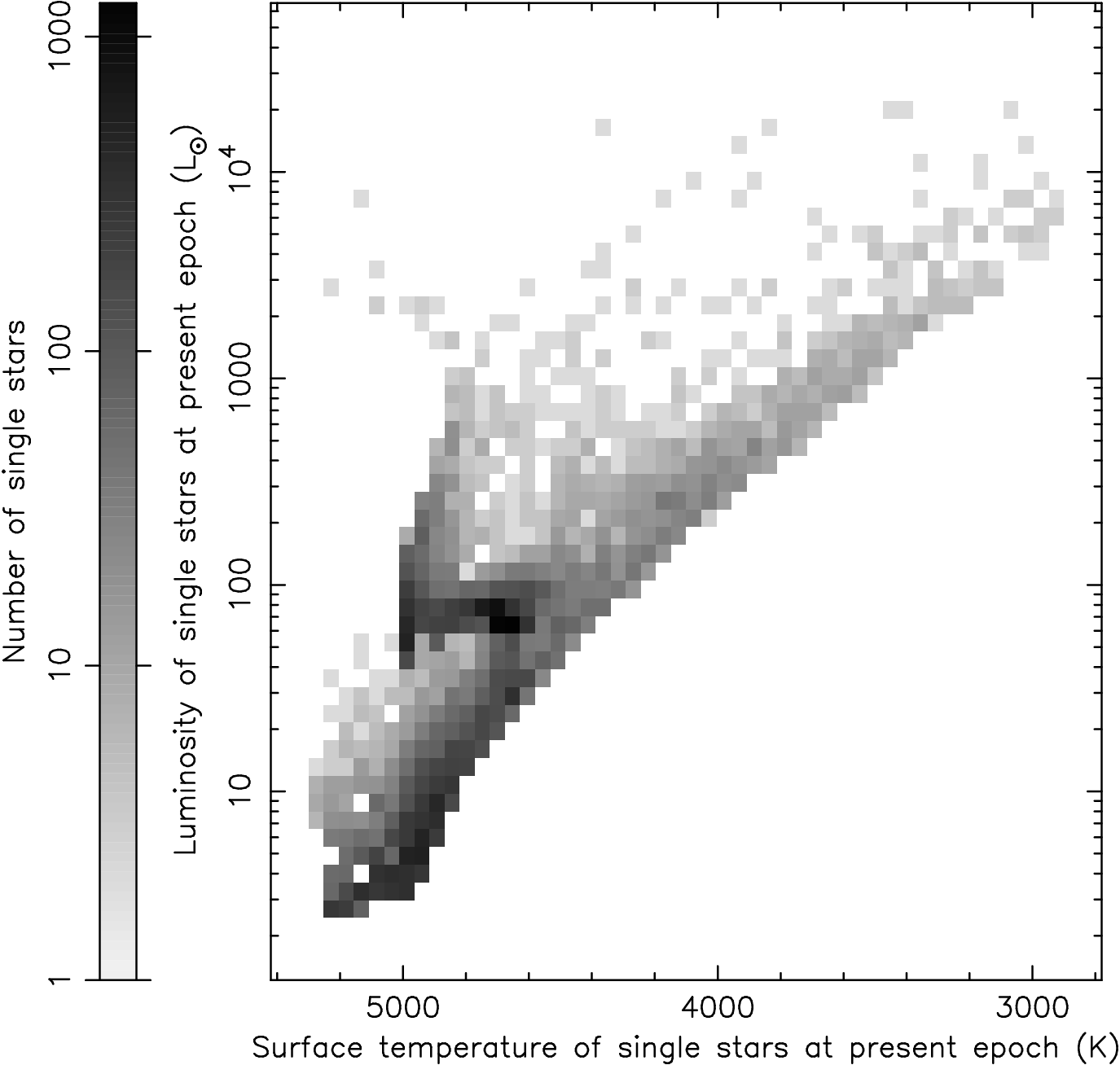}
  \figcaption{
    Two-dimensional distribution of the luminosities and effective temperatures in a comparison model of normal 
    present-day single stars constructed from the same grid of stellar evolution models used for the merged 
    population.  The grey scale represents the total number of merged objects per two-dimensional bin.  
    Temperature increases to the left on the horizontal axis to facilitate comparison with observational HR 
    diagrams.
    \label{fig:hrd_2D_single}
  } 
\end{figure}

In Fig.\,\ref{fig:hrd_2D_single}, we show a present-day, binned, theoretical HR diagram for our normal single-star population. Comparison of 
Fig.\,\ref{fig:hrd_2D_single} with the corresponding plot for merged objects (Fig.\,\ref{fig:hrd_merger}a) indicates little difference between 
the two plots.  This apparent similarity results from the use of our grid of normal single-star models to calculate the structure and evolution 
of the merged objects.  This approximation was necessary because of the 
lack of suitable stellar structure and evolution models for objects that merge during the CE phase.
The degree of error introduced in our predicted luminosities and effective temperatures by using 
non-rotating stellar models is not clear. 

\section{Discussion}
\label{sec:discussion}

We have shown that a distinguishing characteristic feature of a merged stellar population is the high rotational
velocity in comparison with a single-star population. Specifically, rapid rotation in the merged population 
is expected for a fraction of
stars in their giant phase or horizontal-branch phase of evolution. Among observed RGB stars, rapid rotators\footnote{We 
follow the literature in defining rapid rotators as those RGB stars with $v \sin i \geq 10$\,km\,s$^{-1}$.} account for 
1\% to 6\% of the red giant population, depending on the specific observational sample \citep[see][and references 
therein]{car09}. Fig.~\ref{fig:vrot-kms_1D_merger} shows that $\sim$70\% of the projected rotational velocities in our standard merged object population 
are greater than 10\,km\,s$^{-1}$.  Combining this with the number of RGB stars in our merged object and normal single star model populations 
from Table~\ref{tab:dependence}, and assuming a ZAMS binary fraction of 0.5, 
we estimate the fraction of rapid rotators in the RGB population due to CE mergers is $\sim 3.4\%$.   
We note that this estimate neglects RGB stars in binaries whose orbits are wide enough to avoid mass transfer, 
since such binaries are excluded from our population of merged objects.  Inclusion of these binaries increases the overall number of RGB stars, 
which would lower our estimate.  
However, it is also true that tidal interactions between the RGB star and its lower-mass companion can still spin up the giant star in some of these detached binaries. 
Even though such a system retains 
its binarity, the presence of the companion may be masked by the much brighter RGB star and the system could be 
misidentified observationally as a solitary RGB rapid rotator.  At the moment, it is uncertain to what extent such misidentifications are present in the observational 
estimates above.  We also point out that \citet{car09} estimate that rapid rotators produced by the ingestion of planets by RGB stars, which we do not consider, can 
account for $\sim 0.6\%$ of the RGB population.

The subsequent evolution of rapidly rotating RGB stars can lead to the production of rapidly rotating
HB stars as a result of the contraction of stars from the tip of the RGB to the HB when
helium is ignited in the core (see \S\,\ref{sec:results}). 
Although we find theoretical evidence for the rapid rotation of stars in a clump on the red portion of the HB in our HR diagram
(see Fig.\,\ref{fig:hrd_merger}), observational evidence for rapid rotation among red HB field stars is lacking (e.g., \citealp{behr03,dem04}).
However, in a sample of 45 HB field stars, \citet{behr03} found evidence for rapid rotation ($v \sin i 
\gtrsim 30$ km s$^{-1}$) in several blue HB field stars (7500\,K\,$< T <$\,11,500\,K).  He suggests that the underlying
distribution of actual rotational velocities in blue HB field stars may be bimodal, consisting of a slow population ($v \sin i \sim 10$ km s$^{-1}$) 
and a fast population ($v \sin i \sim 30$ km s$^{-1}$), similar to the bimodal distribution found in many globular clusters. 
If a larger sample of HB field stars confirms the possibility of a bimodal distribution in rotational
velocities, the fast-rotating population may result from the mergers studied in this paper. As already
pointed out by \citet{pol08}, the merged population of very rapidly rotating HB stars may
evolve under the action of enhanced stellar wind mass loss due to the reduction of surface gravity
associated with centrifugal effects.  As a result, some very rapidly rotating HB stars may lose a
significant fraction of their envelope, which would cause both a reduction in the rotation rate of
the star because of the associated angular momentum loss, and a blueward movement in the HR diagram.
Such stars may contribute to the rapid rotation of blue HB stars in the field found by \citet{behr03} and,
if sufficient mass is lost, provide an evolutionary channel for the formation of 
single sdB stars.  
\citet{pol08} predict a spectrum of masses ranging from 0.32 
to $\sim 0.7\Msun$ with a strong peak between 0.47 and $0.54\Msun$. This is consistent with 
the small number of observationally determined masses for single sdB stars, the majority of which have masses 
between 0.46 and $0.54\Msun$, but can be as small as $0.39\Msun$ \citep[PG 0911+456;][]{ran07}.  
We further note that, although our models have solar metallicity, if similar rapid rotation and subsequent mass
loss on the HB is found for merger models with lower metallicities, 
this may be relevant to the extended HB morphologies found in globular clusters (e.g., \citealp{mon07}).

Another class of stars that possibly may result from a merger during CE evolution are FK Comae stars.
These stars are rapidly rotating giants with $v \sin i \sim 100$\,km\,s$^{-1}$ and spectral types G and K 
\citep{bs81}.  An origin involving the evolution of a rapidly rotating single MS star is not a viable
one, since the progenitors of FK Comae stars would have rapid rotation and be chromospherically active, 
whereas none have been discovered. On the other hand, the coalescence of evolved contact binaries \citep[see][]
{web76} remains as a viable evolutionary scenario, and we suggest that the merger of non-corotating CE binaries 
may also contribute to this population.  We note, however, that our current model population does not contain 
merged objects of the correct spectral type, although this may be due to our use of non-rotating stellar models 
for these objects. Incorporation of stellar models that include rotation would be especially important in determining the viability
of the merger scenario for FK Comae stars.
  
As a consequence of the merger process, it is likely that there exists a nonspherical distribution of
circumstellar matter surrounding the remnant, either resulting from matter which is not accreted nor ejected
from the system during merger or from an enhanced mass-loss phase from the equatorial region of the rapidly
rotating star.  In the former case, there may exist a short phase (depending on the mass in the
circumstellar envelope) where the interaction of the stellar wind with this disk structure can lead to the
development of a bipolar outflow (where outflow in the equatorial region is impeded by the presence of the
circumstellar disk). Furthermore, it is likely that the mass-loss history of the remnant following the merger event
would be variable, and that the circumstellar matter distribution would reflect its integrated history since
the merger event.  For example, such a temporal history may be responsible for the formation of a torus-like
structure embedded within a continuous outflow, the existence of which has been postulated for the shaping of
planetary nebulae in the generalized interacting-stellar-winds model \citep[see][]{kw82,kw85}, and which
has been inferred from HST observations of the proto-planetary nebula IRAS 17106-3046 \citep[see][]{kw00}.

Such structures are of interest, especially for AGB stars, since observational evidence of axisymmetric
circumstellar envelopes in post-AGB stars and pre-planetary nebulae exists \citep[\textit{e.g.},][]{st98}.  The
mass loss in AGB stars is generally spherically symmetric \citep[\textit{e.g.},][]{bo83,sb93,ol04}, 
although observational evidence of asymmetrical structure exists in the
AGB star V Hya \citep[see][]{hi04}.  Given this state of affairs, the evolutionary state of stars leading
to the transition from a spherical
circumstellar envelope to a nonspherical morphology is of great interest both theoretically and observationally.
The results of our study suggest that the frequency of occurrence of nonspherical circumstellar-envelope
structures resulting from the merger of the components of a binary system during CE evolution is not 
high (at best $\sim 10\%$, assuming all merged AGB stars contain asymmetric envelopes, although it is not 
clear whether mergers between AGB stars and planets would lead to asymmetric envelope structures as well).  
Hence, if the nonspherical structures are not uncommon and a result of binary star interactions, then they may 
reflect the population of detached systems in wide binaries \citep[\textit{e.g.},][]{hmw09} or those remnant 
binary systems that have survived the CE phase resulting in short orbital periods. An extension of our 
population synthesis study to examine the orbital properties of the detached systems and the effect of 
gravitational focusing on the shape of the AGB envelopes will be the subject of a future investigation.

The effect of rotation in the deeper interior layers of the merger remnant may affect the composition of matter
in the surface layers if rotationally induced mixing of matter develops. In such cases, the elemental abundances
in the surface layers may show anomalies compared to those of normal single stars, as matter may be mixed 
from the outer radiative regions of the hydrogen burning shell to the base of the convection zone and, 
hence, to the surface.  The changes in the abundances, however, sensitively depend on
details regarding the magnitude of the diffusion coefficient and the depth of the mixing \citep[see for example,][]{pa06}.

Of particular interest with regard to the nucleosynthetic anomalies at stellar surfaces are the 
Li-rich stars.  These stars are located near the red-giant bump region at luminosities of $\sim 60\Lsun$, 
and in the early AGB phase at luminosities of $\sim 600\Lsun$ \citep[see][]{rl05,kr09}.  
The stars near the red-giant bump may have masses of the order of 
$\sim 1-2.4\Msun$, based on comparison of their luminosities and effective temperatures with 
red-giant evolutionary tracks. A comprehensive discussion of the theoretical models (including the 
merger of a red giant with a Jupiter mass type planet), involving the conversion of $^3$He to $^7$Be, 
the formation of $^7$Li via electron capture \citep{cf71}, and mixing associated with 
an enhanced diffusion coefficient (presumably a result of rapid rotation), is given in \citet{dh04}. 
If the effectiveness of the rotation-induced mixing is related to the location of the
bottom of the surface convection zone relative to the hydrogen burning shell, then such mixing may take 
place in stars above the base of the giant branch, as the mass lying between the base of the convection zone 
and the outer edge of the hydrogen burning shell decreases with increasing evolutionary state. 
Given that the proportion of Li-rich giants among rapid rotators ($v \sin i > 8$\,km\,s$^{-1}$) is $\sim 50\%$ 
\citep[see][]{dp08}, which is considerably greater than the $\sim 2\%$ of Li-rich stars
among slowly rotating stars ($v \sin i < 1$\,km\,s$^{-1}$), it is suggestive that rapid rotation may play an important 
role in forming a significant fraction of the Li-rich giant population.

An additional observation suggesting a possible connection between Li-rich stars and the merged remnants envisioned
here is the relation between the Li abundance and the far-infrared excess in these stars.  In particular, \citet{del96} 
suggest that an enhanced mass loss event may accompany the Li-enrichment phase.  The mass loss resulting from
the phase is expected to be significantly greater (about 2 orders of magnitude) than that for normal Li-poor
giants.  It should be noted that the far-infrared emission may have its origin in a disk-like configuration resulting
from the merging process rather than an expanding dust shell. 

\section{Summary and conclusions}
\label{sec:conclusions}

The population of binary star systems consisting of an RGB or AGB star with a lower-mass 
companion that merge during CE evolution has been explored via a population synthesis technique. 
Using a suite of 116 evolution tracks for   
single stars and including the rotational evolution of the primary components, we have evolved $10^7$  
ZAMS binary systems. The characteristic properties of the merger remnants have been examined as a function of uncertain 
population synthesis input parameters, such as  the initial mass-ratio distribution of the binary population and the efficiency for mass ejection in the  
CE phase, as well as the critical rotational velocity at which enhanced mass loss occurs for the merged object.  Using simplified prescriptions  
for the binary interaction during the CE stage and for the rotational and nuclear evolution of the  
merger remnant, we find that the present-day merged population is primarily 
distinguished by its high rotation speed.  For the total merged population, the median projected rotational velocity is 
16.2$\,$km\,s$^{-1}$, in contrast to 2.3\,km\,s$^{-1}$ for a theoretical population of normal single stars calculated using the 
same stellar models and initial mass function. 
This relative difference is also reflected in individual sub-populations of merged objects (see below), whose peak rotational
velocities are an order of magnitude higher than those in corresponding sub-populations of normal single stars. 

In our present-day model population of merged objects, we include only those objects that are on RGB, HB or AGB.  
We find that these objects constitute between 0.24\% and 0.33\% of the initial population of 10$^7$ ZAMS binaries, depending upon the 
values chosen for the input parameters. In addition, we find
that HB stars are the most populous (57\%), followed by RGB stars (37\%) and then AGB stars (6\%).  The luminosities 
of these sub-groups overlap and lie in the range of $\sim 10 - 100\Lsun$, with the majority of the present-day 
merged population consisting of HB stars with luminosities of $\sim 60\Lsun$, virtually indistinguishable from 
their present-day, normal single-star counterparts 
with the exception of their high rotation speed.  The masses characteristic of the total
population in our standard model are $\lesssim 2\Msun$, with median 
masses for RGB ($1.20\Msun$), HB ($1.35\Msun$) and AGB ($1.34\Msun$) stars that do not significantly differ 
from those for normal single stars on the RGB, but are clearly lower than those for single stars
on the HB and AGB.  The median mass of the total population is about $0.05\,M_\odot$ higher for our
standard merger population than for a normal single-star population, which may seem to contradict the results
per sub-population, but can be explained by the different RGB versus HB ratios between merged objects and normal 
single stars (see \S\,\ref{sec:single}).
These quantitative features of the present-day merged population are found to be relatively insensitive to the
population synthesis input parameters.

The results of our population synthesis study have been discussed in terms of possible observational counterparts  
either directly involving the high rotational velocity of the component or indirectly, via the effect of rotation on  
the amount and distribution of circumstellar matter. Additional probes potentially indicating the occurrence of such   
interactions could be provided by compositional signatures on the stellar surface.  
As a specific application of our CE merger model, we have considered  
the possibility that rotationally enhanced mass loss may be sufficient to significantly deplete low- or 
intermediate-mass hydrogen-rich envelopes of rapidly rotating HB stars to the extent that their effective temperatures 
are increased sufficiently to produce sdB stars. Such a merger scenario can provide an additional evolutionary channel  
for the formation of single sdB stars \citep{sok00,sok07,pol08}, complementary to the double 
helium-WD merger channel proposed earlier by \citet{han03}.   
 
Indirect effects associated with the merger and rapid rotation, coupled with the evolution of the merged object to   
its later phases, can lead to deviations from spherical symmetry in the distribution of circumstellar matter.  Such  
structures can be potentially probed via their emission in the radio \citep[\textit{e.g.},][]{hi04} or infrared wavelength 
region, or via the scattering of optical light \citep[\textit{e.g.},][]{mh06}. These structures may result from inherent 
asymmetries present in the merger process itself, or result after the merger has already occurred, reflecting asymmetric
mass loss from the stellar surface
or the interaction of a spherically symmetric stellar wind with a nonspherical distribution of circumstellar matter. 
Additional observational effects may be a consequence  
of rotation in the deep interior, possibly resulting from rotationally induced compositional mixing leading to the production  
of nucleosynthetic anomalies at the stellar surface. In this regard, the merger scenario may provide an 
additional evolutionary channel for producing Li-rich RGB stars.

To gain insight into the outcome of the merger scenario, our investigation has been primarily exploratory in nature.  As a 
result, the study has been based on a number of simplifying assumptions. 
Foremost among them is the treatment of the structure and
evolution of a merged object as equivalent to a single, non-rotating star of the same total mass and core mass. The
binary interaction involving the evolution into and during the CE phase is also uncertain, as the efficiency of mass
ejection has been assumed to be a constant, independent of evolutionary phase.  Mass loss prior to and after the CE phase
has been based on rates inferred for normal single stars, which are likely to have much smaller rotational velocities.
Finally, it is likely that the mass loss via a stellar wind, and hence angular momentum loss from
the system, would be affected by the spin-orbit tidal coupling and magnetic braking.
  
To improve on the numerical modeling, future theoretical work is indicated, especially in including neglected effects. For 
example, the use of a grid of stellar models resulting from the merger process \citep[see for example,][]{gleb08a} would be  
particularly illuminating, not only in describing the structure and properties of the merged remnant, but also in determining 
their evolutionary characteristics as a function of age. This would be most useful for our large population of merged 
objects on the RGB and HB.
Our work also provides valuable guidance by better defining 
the region of parameter space where detailed multi-dimensional hydrodynamical calculations are highly desirable.  
We find that mergers during CE evolution likely happen in binaries characterized by primaries with masses of $\sim 1-2\Msun$ 
that fill their Roche lobes on the lower half of the RGB at orbital periods between $\sim 1$ and 30 days.
Our study also points to the phases of evolution where hydrodynamical or magnetohydrodynamical calculations are needed to 
explore the influence of rotation and/or magnetic fields in the mixing of matter and angular momentum 
from the the deep interior (e.g., regions that have undergone nuclear processing) to the stellar surface.  
Incorporation of mixing prescriptions, based on the above 
studies, into evolutionary calculations of a merged remnant could provide quantitative estimates of the influence 
of rotation on the surface composition.  Of special interest would be numerical modeling of the rotating remnant 
during a helium flash, as it may provide a compositional probe of the possible mixing of matter between the helium core and 
hydrogen-rich envelope for stars that are now on the HB.
Finally, an extension of our models to examine the properties of binaries that merge 
in the MS phase of evolution is of   
interest, as they may have applications to highly luminous transient events, perhaps similar to that observed in V 838 Mon   
\citep[see for example,][]{st06}.
  
In addition to the fundamental theoretical work, future observational programs aimed at detecting the effects of 
rapid rotation using ground-based optical interferometers could be fruitful in the determination of the shapes of  
giant stars, where temperature variations of the surface may be probed \citep[\textit{e.g.},][]{zhao09}.  
A program of highly accurate photometric studies can be
considered, enabling one to detect differences in luminosity and colors of rapidly rotating stars in comparison to their single-star
counterparts. For example, observational evidence exists that magnetically-active, rapidly rotating, low-mass MS stars in
eclipsing binaries are characterized by effective temperatures that are lower than, radii that are larger than, and luminosities
that are approximately equal to those predicted from corresponding theoretical stellar models of the same mass \citep[\textit{e.g.},][]{tor09}.
The same stellar models satisfactorily match observations of similar-mass stars that are rotating more slowly in wide binary orbits.
\citet{tor06}, \citet{lop07}, \citet{mor08} and others have shown that the source of these differences is the magnetic activity in
the rapidly rotating stars.  Whether the radii, effective temperatures and luminosities in rapidly rotating giant or HB stars would
be similarly affected is an interesting and unresolved question.
Finally, studies on the asymmetries in the circumstellar envelope of stars in the AGB phase with upcoming 
facilities (\textit{e.g.}, ALMA) will be especially useful in providing constraints on the stellar evolutionary phases where  
asymmetries develop.  Such studies will be very important for distinguishing the contributions of the merged population 
from the existing binary population, thereby potentially providing further constraints on uncertain 
population synthesis input parameters  
and possible shaping mechanisms for the origin of the asymmetries seen to be prevalent in the post-AGB and proto-planetary   
nebula phase.

\acknowledgements{
  We warmly thank P.P.\ Eggleton for making his binary-evolution code available 
  to us.  M.P.\ acknowledges funding from the Wisconsin Space Grant Consortium and NSF grant AST-0607111, sub-award 1-2008, 
  to Marquette University.  MvdS acknowledges support from NSF CAREER Award AST-0449558 to Northwestern University and 
  a CITA National Fellowship to the University of Alberta.  R.T.\ and B.W.\ acknowledge support from NSF AST-0703950 and 
  NASA BEFS NNG06GH87G, respectively, to Northwestern University.  Lastly, we thank the referee, Dr.\ Evert Glebbeek, 
  for a thorough reading of our manuscript and many useful comments that improved the paper.
}


\begin{thebibliography}{}

\bibitem [Abt(1983)] {abt83}
Abt, H.A. 1983, \araa, 21, 343
\bibitem [Baraffe et al.(1998)] {bar98}
Baraffe, I., Chabrier, G., Allard, F., \& Hauschildt, P. 1998, \aap, 337, 403
\bibitem [Baraffe et al.(2003)] {bar03}
Baraffe et al. 2003, \aap, 402, 701
\bibitem[Behr(2003)]{behr03}
Behr, B. B. 2003, \apjs, 149, 101
\bibitem [Bopp \& Stencel(1981)]{bs81} 
Bopp, B. W., \& Stencel, R. E. 1981, \apj, 247, L131
\bibitem [Bowers et al.(1983)]{bo83} 
Bowers, B. F, Johnston, K. J., \& Spencer J. H. 1983, \apj, 274, 733
\bibitem [Busso et al.(2007)]{bus07}
Busso et al. 2007, \aap, 47, 105
\bibitem [Cameron \& Fowler(1971)]{cf71} 
Cameron, A. G. W., \& Fowler, W. A. 1971, \apj, 164, 111
\bibitem [Carlberg, Majewski, \& Arras(2009)]{car09}
Carlberg, J.K., Majewski, S.R., \& Arras, P. 2009, \apj, 700, 832
\bibitem [Catelan(2007)] {cat07}
Catelan, M. 2007, in New Quests in Stellar Astrophysics II: The Ultraviolet Properties of Evolved Stellar Populations, 
ed. M. Chavez, E. Bertone, D. Rosa-Gonzalez, \& L. H. Rodriguez-Merino, (New York: Springer), 175
\bibitem [Chabrier \& Baraffe(1997)] {cb97}
Chabrier, G. \& Baraffe, I. 1997, \aap, 327, 1039
\bibitem [Chabrier \& Baraffe(2000)] {cb00}
Chabrier, G. \& Baraffe, I. 2000, \araa, 38, 337
\bibitem [Chabrier et al.(2000)] {cha00}
Chabrier, G., Baraffe, I., Allard, F., \& Hauschildt, P.H. 2000, \apj, 542, 464
\bibitem [Darwin(1879)] {dar79}
Darwin, G.H. 1879, Proc. Roy. Soc., 29, 168
\bibitem[de Jager et al.(1988)]{1988A&AS...72..259D}
{de Jager}, C., {Nieuwenhuijzen}, H., \& {van der Hucht}, K.~A.\ 1988, A\&AS, 72, 259 
\bibitem [de la Reza et al.(1996)]{del96} 
De la Reza, R., Drake, N. A., \& Da Silva, L. 1996, \apj, 456, L115
\bibitem [de Medeiros(2004)]{dem04} 
de Medeiros, J. R. 2004, in Stellar Rotation, IAU Symposium No. 215, ed. A. Maeder \& P. Eenens
(San Francisco: Astronomical Society of the Pacific), p. 144
\bibitem [Denissenkov \& Herwig(2004)]{dh04} 
Denissenkov, P. A., \& Herwig F. 2004, \apj, 612, 1081
\bibitem [Denissenkov \& Pinsonneault(2008)]{dp08} 
Denissenkov, P. A., \& Pinsonneault, M. 2008, \apj, 684, 626
\bibitem [Dorman et al.(1993)]{dor93}
Dorman, B., Rood, R.T., \& O'Connell, R.W. 1993, \apj, 419, 596
\bibitem [Duquennoy \& Mayor(1991)] {duq91}
Duquennoy, A., \& Mayor, M. 1991, \aap, 248, 485
\bibitem[{{Eggleton}(1971)}]{1971MNRAS.151..351E}
{Eggleton}, P.~P. 1971, MNRAS, 151, 351
\bibitem[{{Eggleton}(1972)}]{1972MNRAS.156..361E}
{Eggleton}, P.~P. 1972, MNRAS, 156, 361
\bibitem[Fryer \& Woosley(1998)]{fry98}
Fryer, C. L., \& Woosley, S. E. 1998, \apj, 502, L9
\bibitem[Fryer \& Heger(2005)]{fry05}
Fryer, C. L., \& Heger, A. 2005, \apj, 623, 302
\bibitem [Goldberg et al.(2003)] {gol03}
{Glebbeek}, E., \& {Pols}, O.~R. 2008, \aap, 488, 1017
\bibitem [Glebbeek \& Pols(2008)] {gleb08a}
Goldberg, D., Mazeh, T., \& Latham, D.W. 2003, \apj, 591, 397
\bibitem [Green, Liebert, \& Saffer(2001)]{gre01}
Green, E. M., Liebert, J., \& Saffer, R. A. 2001, in  in 12th European Workshop on White Dwarfs, ASP Conf. Ser., Vol. 226, ed. J. L. Provencal, H. L. Shipman, J. MacDonald, and S. Goodchild, (San Francisco: ASP), 192
\bibitem [Han et al.(2002)] {han02}
Han, Z., Podsiadlowski, Ph., Maxted, P.F.L., Marsh, T.R., \& Ivanova, N. 2002, \mnras, 336, 449
\bibitem [Han et al.(2003)] {han03}
Han, Z., Podsiadlowski, Ph., Maxted, P.F.L., \& Marsh, T.R. 2003, \mnras, 341, 669
\bibitem [Han et al.(2007)] {han07}
Han, Z., Podsiadlowski, Ph., \& Linas-Gray, 2007, \mnras, 380, 1098
\bibitem [Heber(2008)]{heb08}
Heber, U. 2008, in XXI Century Challenges for Stellar Evolution, ed. S. Cassisi \& M. Salaris, Mem. Soc. Astr. It., Vol. 79, 375
\bibitem [Hirano et al.(2004)]{hi04} 
Hirano, N., Shinnaga, H., Dinh-V., T., Fong, D., Keto, E., Patel, N., Qi, C., Young, K., Zhang, Q, \& Zhao, J.\ 2004, \apj, 616, L43)
\bibitem [Hjellming \& Taam(1991)] {hje91}
Hjellming, M.S., \& Taam, R.E., 1991, \apj, 370, 709
\bibitem [Huggins et al.(2009)]{hmw09} 
Huggins, P. J., Mauron, N., \& Wirth, E. A. 2009, \mnras, 396, 1805
\bibitem [Hurley et al.(2000)] {hur00}
Hurley, J.R., Tout, C.A., \& Pols, O.R. 2000, \mnras, 315, 543
\bibitem [Hurley et al.(2002)] {hur02}
Hurley, J.R., Tout, C.A., \& Pols, O.R. 2002, \mnras, 329, 897
\bibitem [Iben \& Livio(1993)] {ibe93}
Iben, I., Jr., \& Livio, M. 1993, \pasp, 105, 1373
\bibitem [Iben \& Tutukov(1985)] {ibe85}
Iben, I., Jr, \& Tutukov, A.V. 1985, \apjs, 58, 661
\bibitem [Ivanova \& Podsiadlowski(2002)]{ip02}
Ivanova, N., \& Podsiadlowski, Ph., 2002, \apss, 281, 191
\bibitem [Ivanova, Podsiadlowski, \& Spruit(2002)]{iva02}
Ivanova, N., Podsiadlowski, Ph., \& Spruit, H. 2002, \mnras, 334, 819
\bibitem [Izzard, Jeffery, \& Lattanzio(2007)]{izz07}
Izzard, R.G., Jeffery, C.S., \& Lattanzio, J. 2007, \aap, 470, 661
\bibitem [Kahn \& West(1985)]{kw85} 
Kahn, F. D. \& West, K. A. 1985, \mnras, 212, 837
\bibitem [Kumar \& Reddy(2009)]{kr09} 
Kumar, Y.~B., \& Reddy, B.~E. 2009, \apj, 703, L46
\bibitem [Kwok(1982)]{kw82} 
Kwok, S. 1982, \apj, 258, 280
\bibitem [Kwok et al.(2000)]{kw00} 
Kwok, S., Hvrinak, B., \& Su, K. Y. L. 2000, \apj, 544, L149
\bibitem [Lisker et al.(2005)]{lis05}
Lisker et al. 2005, \aap, 430, 223
\bibitem [Lombardi et al.(2002)]{lom02}
Lombardi et al. 2002, \apj, 568, 939
\bibitem [L\'{o}pez-Morales(2007)]{lop07}
L\'{o}pez-Morales M. 2007, \apj, 660, 732
\bibitem [MacLaurin(1742)]{maclaurin} 
MacLaurin, C.\ 1742, A treatise of fluxions, Edinburgh: Ruddimans; Cf.\ Lang, K.~R.\ 1999, Astrophysical formulae, New York: Springer
\bibitem [Mauron \& Huggins(2006)]{mh06} 
Mauron, N. \& Huggins, P. J. 2006, \aap, 452, 257
\bibitem [Maxted et al.(2001)]{max01}
Maxted, P.F.L., Heber, U., Marsh, T.R., \& North, R.C. 2001, \mnras, 326, 1391
\bibitem [Mazeh et al.(1992)] {maz92}
Mazeh, T., Goldberg, D., Duquennoy, A., \& Mayor, M. 1992, \apj, 401, 265
\bibitem[Middleditch(2004)]{mid04}
Middleditch, J. 2004, \apj, 601, L167
\bibitem [Miller \& Scalo(1979)] {mil79}
Miller, G.E., \& Scalo, J.M. 1979, \apjs, 41, 513
\bibitem [Moni Biden et al.(2008)]{mon07}
Moni Biden et al. 2008, in Hot Subdwarf Stars and Related Objects, ASP Conf. Ser., Vol. 392, ed. U. Heber, S. Jeffery, \&
R. Napiwotski (San Francisco: ASP), 27
\bibitem [Moni Biden et al.(2008)]{mon08}
Moni Biden, C., Catelan, M., \& Altmann, M. 2008, \aap, 480, L1
\bibitem [Morales et al.(2008)] {mor08}
Morales J. C., Ribas I., Jordi C. 2008, \aap, 478, 507
\bibitem [Morales-Rueda et al.(2006)] {mor06}
Morales-Rueda et al. 2006, Balt. Astr., 15, 187
\bibitem [Napiwotzki et al.(2004)]{nap04}
Napiwotzki et al. 2004, \apss, 291, 321
\bibitem [Olofsson(2004)]{ol04} Olofsson, H. 2004, in IAU Symp. 191, Asymptotic Giant Branch Stars, ed. T. le Bertre, A. Lebre, \&
C. Waelkens (San Franciso: ASP), 325
\bibitem [Palacios et al.(2006)]{pa06} 
Palacios et al. 2006, \aap, 453, 261
\bibitem[Podsiadlowski, Joss, \& Rappaport(1990)]{pod90}
Podsiadlowski, Ph., Joss, P. C., \& Rappaport, S. 1990, \aap, 227, L9
\bibitem[Podsiadlowski(2001)]{pod01}
Podsiadlowski, Ph. 2001, in Evolution of Binary and Multiple Star Systems, ASP Conf. Ser., Vol. 229, ed. Ph. Podsiadlowski, 
S. Rappaport, A. R. King, F. D'Antona, \& L. Burderi, (San Francisco: ASP), 239
\bibitem [Politano(1996)] {pol96}
Politano, M. 1996, \apj, 465, 338
\bibitem [Politano \& Weiler(2007)] {pw07}
Politano, M., \& Weiler, K.P. 2007, \apj, 665, 663
\bibitem [Politano et al.(2008)] {pol08}
Politano, M., Taam, R. E. , van der Sluys, M., \& Willems, B.. 2008, \apj, 687, L99
\bibitem[{{Pols} {et~al.}(1995){Pols}, {Tout}, {Eggleton}, \& {Han}}]{1995MNRAS.274..964P}
{Pols}, O.~R., {Tout}, C.~A., {Eggleton}, P.~P., \& {Han}, Z. 1995, MNRAS, 274, 964
\bibitem [Randall et al.(2007)]{ran07}
Randall et al. 2007, \aap, 476, 1317
\bibitem [Reddy \& Lambert(2005)]{rl05} 
Reddy, B. E., \& Lambert, D. L. 2005, \aj, 129, 2831
\bibitem [Reed \& Stiening(2004)]{ree04}
Reed, M.D., \& Stiening, R. 2004, \pasp, 116, 506
\bibitem [Reimers(1975)] {rei75}
Reimers, D. 1975, Mem. Soc. Roy. Sci. Liege, 6th, ser., 6, 369
\bibitem[Ricker \& Taam(2008)]{rick08}
Ricker, P. M., \& Taam, R. E. 2008, \apj, 672, L41
\bibitem [Sahai \& Bieging(1993)]{sb93} 
Sahai, R., \& Bieging, J. H. 1993, \aj, 105, 595
\bibitem [Sahai \& Trauger(1998)]{st98} 
Sahai, R., \& Trauger, J. T. 1998, \aj, 116, 1357
\bibitem[Sandquist, Taam, \& Burkert(2000)]{san00}
Sandquist, E. L., Taam, R. E., \& Burkert, A. 2000, \apj, 533, 984
\bibitem[Sandquist et al.(1998)]{san98}
Sandquist et al. 2000, \apj, 500, 909
\bibitem[Siess \& Livio(1999a)]{sie99a}
Siess, L., \& Livio, M. 1999a, \mnras, 304, 925
\bibitem[Siess \& Livio(1999b)]{sie99b}
Siess, L., \& Livio, M. 1999b, \mnras, 308, 1133
\bibitem [Sills et al.(1997)]{sil97}
Sills et al. 1997, \apj, 487, 290
\bibitem [Sills et al.(2001)]{sil01}
Sills et al. 2001, \apj, 548, 323
\bibitem [Soker \& Harpaz(2000)]{sok00}
Soker, N., \& Harpaz, A. 2000, \mnras, 317, 861
\bibitem [Soker \& Tylenda (2006)]{st06} 
Soker, N., \& Tylenda, R. 2006, \mnras, 373, 733
\bibitem [Soker \& Harpaz(2007)]{sok07}
Soker, N., \& Harpaz, A. 2007, \apj, 660, 699
\bibitem [Sweigart(1997)] {sw97}
Sweigart, A. V. 1997, \apj, 474, L23
\bibitem [Taam \& Sandquist(2000)] {taa00}
Taam, R.E., \& Sandquist, E.L., 2000, \araa, 38, 113
\bibitem [Torres et al.(2006)]{tor06}
Torres G., Lacy C. H., Marschall L. A., Sheets H. A., \& Mader J. A. 2006, \apj, 640, 1018
\bibitem [Torres et al.(2009)]{tor09}
Torres, G., Andersen, J., \& Gimenez, A. 2009, arXiv:0908.2624v1
\bibitem [Tutukov \& Yungleson(1979)] {tut79}
Tutukov, A.V., \& Yungelson, L.R. 1979, in Mass Loss and Evolution of O-Type Stars, ed. P.S. Conti \& W.H. de Loore (Dordrecht:  Reidel), 216
\bibitem [Tutukov \& Yungleson(1990)] {tut90}
Tutukov, A.V., \& Yungelson, L.R. 1990, Sov. Astr., 34, 57
\bibitem [Tutukov \& Yungleson(2005)] {tut05}
Tutukov, A.V., \& Yungelson, L.R. 2005, Astr. Rep., 49, 871
\bibitem[{{van der Sluys} {et~al.}(2006){van der Sluys}, {Verbunt}, \& {Pols}}]{2006A&A...460..209V}
{van der Sluys}, M.~V. and {Verbunt}, F. and {Pols}, O.~R. 2006, A\&A, 460, 209
\bibitem [Webbink(1976)]{web76} 
Webbink, R. F. 1976, \apj, 209, 829
\bibitem [Willems et al.(2005)] {wil05}
Willems, B., Kolb, U., Sandquist, E.L., Taam, R.E., \& Dubus, G. 2005, \apj, 635, 1263
\bibitem[{Yakut \& {Eggleton}(2005)}]{2005ApJ...629..1055Y}
{Yakut}, K. \& {Eggleton}, P.~P. 2005, ApJ, 629, 1055
\bibitem [Zhao et al.(2009)] {zhao09} 
Zhao, M. et al. 2009, \apj, 701, 209
\end{thebibliography}
\end{document}